\documentclass[twocolumn, nofootinbib, notitlepage,superscriptaddress, secnumarabic, aps]{revtex4-1}
\usepackage{bm, bbm, bbding}
\usepackage{amssymb,amsmath,amsfonts,mathrsfs,graphicx,xcolor,units,yfonts}
\usepackage{txfonts,dsfont}
\usepackage[toc,page,title,titletoc,header]{appendix}
\usepackage{setspace, framed, hyperref}
\usepackage[novbox]{pdfsync}
\hypersetup{
  colorlinks=true,        
  linkcolor=blue,         
  citecolor=blue,       
  urlcolor=black  }

\begin{document}

\title{Lovelock-Brans-Dicke gravity}

\author{David Wenjie Tian}%
\email{wtian@mun.ca}
\affiliation{Faculty of Science,  Memorial University, St. John's, Newfoundland, Canada, A1C 5S7}
\author{Ivan Booth}%
\email{ibooth@mun.ca}
\affiliation{Department of Mathematics and Statistics, Memorial University, St. John's,  Newfoundland, Canada, A1C 5S7}

\begin{abstract}
According to Lovelock's theorem, the Hilbert-Einstein and the Lovelock actions are indistinguishable from their field equations. However, they have different scalar-tensor counterparts, which correspond to the Brans-Dicke and the \emph{Lovelock-Brans-Dicke} (LBD) gravities, respectively. In this paper the LBD model of alternative gravity with the Lagrangian density $\mathscr{L}_{\text{LBD}}=\frac{1}{16\pi}\left[\phiup\left( R
+\frac{a}{\sqrt{-g}}{}^*RR + b\mathcal{G}\right)-\frac{\omega_{\text{L}}}{\phiup}\nabla_\alpha \phiup \nabla^\alpha\phiup \right]$ is developed, where ${}^*RR$ and $\mathcal{G}$ respectively denote the topological Chern-Pontryagin and Gauss-Bonnet invariants. The field equation, the kinematical and dynamical wave equations, and the constraint from energy-momentum conservation are all derived. It is shown that, the LBD gravity reduces to general relativity in the limit $\omega_{\text{L}}\to\infty$ unless the ``topological balance condition'' holds, and in vacuum it can be conformally transformed into the dynamical Chern-Simons gravity and the generalized Gauss-Bonnet dark energy with Horndeski-like or Galileon-like kinetics. Moreover, the LBD gravity allows for the late-time cosmic acceleration without dark energy. Finally, the LBD gravity is generalized into the Lovelock-scalar-tensor gravity, and its equivalence to fourth-order modified gravities is established. It is also emphasized that the standard expressions for the contributions of generalized Gauss-Bonnet dependence can be further simplified. \\

\noindent Key words: Lovelock's theorem, topological effects, modified gravity
\end{abstract}

\maketitle


\section{Introduction}\label{Section Introduction}

As an alternative to the various models of dark energy
with large negative pressure that violates the standard energy conditions,
the accelerated expansion of the Universe has inspired the reconsideration of relativistic gravity and modifications of general relativity (GR), which can explain the cosmic acceleration and reconstruct the entire expansion history without dark energy.


Such alternative and modified gravities actually encode the possible  ways to go beyond Lovelock's theorem and its necessary conditions \cite{Lovelock theorem I}, which limit the
second-order field equation in four dimensions to $R_{\mu\nu}-Rg_{\mu\nu}/2+\Lambda g_{\mu\nu}=8\pi G T_{\mu\nu}^{\text{(m)}}$, i.e. Einstein's equation supplemented by the cosmological constant $\Lambda$. These directions can allow for, for example, fourth and even higher order gravitational field equations \cite{Carroll R+f(R Rc2 Rm2)+2kLm, GaussBonnet first model R/2k+f(G), GaussBonnet second model f(R G)+Lm, fR review}, more than four spacetime dimensions \cite{Lovelock gravity generic, Clifton MG Lovelock}, extensions of pure pseudo-Riemannian geometry and metric gravity \cite{Clifton MG Lovelock, Teleparalel}, extra physical degrees of freedom \cite{Brans Dicke, Scalar Tensor Theory, Chern-Simons 1, GaussBonnet dark energy}, and nonminimal curvature-matter couplings \cite{Nonminimal coupling, AA Tian-Booth Paper}. From a variational approach, these violations manifest themselves as different modifications of the Hilbert-Einstein action, such as extra curvature invariants, scalar fields, and non-Riemannian geometric variables.

For the Lovelock action in Lovelock's theorem and the Hilbert-Einstein-$\Lambda$ action, it is well known that  they yield the same field equation and thus are indistinguishable by their gravitational effects. When reconsidering Lovelock's theorem, we cannot help but ask whether the effects of these two actions are really the same in all possible aspects. Is there any way for the two topological sources in the Lovelock action to show nontrivial consequences?
As a possible answer to this question, we propose the Lovelock-Brans-Dicke gravity.

This paper is organized as follows. In Sec.~\ref{Sec Lovelock-Brans-Dicke gravity}, the Lovelock-Brans-Dicke gravity is introduced based on Lovelock's theorem, and its gravitational and wave equations are derived in Sec.~\ref{Gravitational and wave equations LBD}.
Section~\ref{Sec The omega to infty limit and GR} studies the behaviors at the infinite-Lovelock-parameter limit $\omega_{\text{L}}\to\infty$, and Sec.~\ref{Sec Constraints from energy-momentum conservation} derives the constraint from energy-momentum conservation. Section~\ref{Sec Conformal transformations} shows that in vacuum the Lovelock-Brans-Dicke gravity can be conformally transformed into the dynamical Chern-Simon gravity and the generalized Gauss-Bonnet dark energy with Horndeski-like or Galileon-like kinetics.
Then the possibility of realizing the acceleration phase for the late-time Universe is discussed in Sec.~\ref{Cosmological applications}.
Finally, in Sec.~\ref{Lovelock-scalar-tensor gravity} the Lovelock-Brans-Dicke theory is extended to the Lovelock-scalar-tensor gravity, and its equivalence to fourth-order modified gravities is analyzed.
Throughout this paper,  we adopt the sign conventions $\Gamma^\alpha_{\beta\gamma}=\Gamma^\alpha_{\,\;\;\beta\gamma}$,
$R^{\alpha}_{\;\;\beta\gamma\delta}=\partial_\gamma
\Gamma^\alpha_{\delta\beta}-\partial_\delta \Gamma^\alpha_{\gamma\beta}\cdots$ and  $R_{\mu\nu}=R^\alpha_{\;\;\mu\alpha\nu}$ with the metric signature $(-,+++)$.


\section{Lovelock-Brans-Dicke action}\label{Sec Lovelock-Brans-Dicke gravity}

An algebraic Riemannian invariant $\widetilde{\mathcal{R}}=\widetilde{\mathcal{R}}
\left(g_{\alpha\beta}\,,R_{\alpha\mu\beta\nu}\right)$ in the action $\int d^4x\sqrt{-g}\,\widetilde{\mathcal{R}}$ generally leads to fourth-order gravitational field equations by the variational derivative
\begin{equation}
\frac{\delta \left(\!\!\sqrt{-g}\widetilde{\mathcal{R}}\right)}{\delta g^{\mu\nu}}=
\frac{ \partial \left(\!\!\sqrt{-g}\widetilde{\mathcal{R}}\right)}{\partial g^{\mu\nu}}-\partial_\alpha \frac{ \partial \left(\!\!\sqrt{-g}\widetilde{\mathcal{R}}\right)}{\partial (\partial_\alpha g^{\mu\nu})}
+\partial_\alpha \partial_\beta\frac{\partial \left(\!\!\sqrt{-g}\widetilde{\mathcal{R}}\right)}
{\partial (\partial_\alpha \partial_\beta g^{\mu\nu})}.
\end{equation}
Lovelock found out that in \emph{four} dimensions the most general action leading to second-order field equations is \cite{Lovelock theorem I}
\begin{eqnarray}\label{Lanczos-Lovelock action}
\mathcal{S}&&=\int d^4x \sqrt{-g}\,\mathscr{L} +\mathcal{S}_m\quad\mbox{with}\\
\mathscr{L}&&= \frac{1}{16\pi G}\left( R-2\Lambda
+\frac{a}{2\sqrt{-g}}\epsilon_{\alpha\beta\mu\nu}
R^{\mu\nu}_{\;\;\;\;\,\gamma\delta}R^{\alpha\beta\gamma\delta}+b\mathcal{G}\right),\nonumber
\end{eqnarray}
where $\Lambda$ is the cosmological constant, $\{a \,,b\}$ are dimensional coupling constants, and without any loss of generality we have set the coefficient of $R$ equal to one. Also,  $\epsilon_{\alpha\beta\mu\nu}$ refers to the totally antisymmetric Levi-Civita pseudotensor with $\epsilon_{0123}=\sqrt{-g}$, $\epsilon^{0123}=\frac{1}{\sqrt{-g}}$, and $\{\epsilon_{\alpha\beta\mu\nu}$, $\epsilon^{\alpha\beta\mu\nu}\}$ can be  obtained from each other by raising or lowering the indices with the metric tensor. In Eq.(\ref{Lanczos-Lovelock action}), $\epsilon_{\alpha\beta\mu\nu} R^{\mu\nu}_{\;\;\;\;\,\gamma\delta}R^{\alpha\beta\gamma\delta}$ and $\mathcal{G}$ respectively refer to the Chern-Pontryagin density and the Gauss-Bonnet invariant, with
\begin{equation}
\mathcal{G}\,\coloneqq\,R^2-4R_{\mu\nu}R^{\mu\nu}
+R_{\mu\alpha\nu\beta}R^{\mu\alpha\nu\beta}.
\end{equation}

The variational derivatives
$ \delta(\epsilon_{\alpha\beta\mu\nu} R^{\mu\nu}_{\;\;\;\;\,\gamma\delta}
R^{\alpha\beta\gamma\delta} )/\delta g^{\mu\nu}$ and $\delta (\!\sqrt{-g}\mathcal{G})/\delta g^{\mu\nu} $
yield total derivatives which serve as boundary terms in varying the full action Eq.(\ref{Lanczos-Lovelock action}). The Chern-Pontryagin scalar $\epsilon_{\alpha\beta\mu\nu} R^{\mu\nu}_{\;\;\;\;\,\gamma\delta}R^{\alpha\beta\gamma\delta}$ is proportional to the divergence of the topological Chern-Simons four-current $K^\mu$  \cite{Chern-Simons 1},
\begin{equation}\label{CP current}
\begin{split}
&\epsilon_{\alpha\beta\mu\nu} R^{\mu\nu}_{\;\;\;\;\,\gamma\delta}R^{\alpha\beta\gamma\delta} = -8\,\partial_\mu K^\mu \quad\text{with}\\
&K^\mu = \epsilon^{\mu\alpha\beta\gamma}
\left(\frac{1}{2}\Gamma^\xi_{\alpha\tau}\partial_\beta\Gamma^\tau_{\gamma\xi}
+\frac{1}{3} \Gamma^\xi_{\alpha\tau}\Gamma^\tau_{\beta\eta}\Gamma^\eta_{\gamma\xi}  \right),
\end{split}
\end{equation}
and similarly, the topological current for the Gauss-Bonnet invariant is (see Refs.\cite{Chern Simons scalar, Gauss-Bonnet current 1} for earlier discussion and Ref.\cite{Gauss-Bonnet current 2} for further clarification)
\begin{equation}\label{GB current}
\begin{split}
&\sqrt{-g}\, \mathcal{G}=-\partial_\mu J^\mu\quad\mbox{with}\\
 J^\mu = &\sqrt{-g}\, \epsilon^{\mu\alpha\beta\gamma}\epsilon_{\rho\sigma}^{\;\;\;\;\xi\zeta}\Gamma^\rho_{\xi\alpha}
 \left(\frac{1}{2}R^\sigma_{\;\;\zeta\beta\gamma}
 -\frac{1}{3}\Gamma^\sigma_{\lambda\beta}\Gamma^\lambda_{\zeta\gamma}
  \right).
\end{split}
\end{equation}
Hence, the covariant densities $\epsilon_{\alpha\beta\mu\nu} R^{\mu\nu}_{\;\;\;\;\,\gamma\delta}R^{\alpha\beta\gamma\delta}$ and $\sqrt{-g}\mathcal{G}$ in Eq.(\ref{Lanczos-Lovelock action}) make no contribution to the field equation $\delta \mathcal{S}/\delta g^{\mu\nu}=0$.

According to Lovelock's theorem,
one cannot tell whether Einstein's equation $R_{\mu\nu}-\frac{1}{2}R
g_{\mu\nu}=8\pi G T_{\mu\nu}^{\text{(m)}}$ comes from the customary Hilbert-Einstein action
\begin{equation}\label{Preparations Hilbert action}
\mathcal{S}_{\text{HE}}\,=\,\frac{1}{16\pi G}\int d^4x \sqrt{-g}\: R + \mathcal{S}_m\,,
\end{equation}
or from the induced Lovelock action\footnote{Note that not  to confuse with the more common  ``Lovelock action''  for the topological generalizations of the Hilbert-Einstein action to  generic $N$ dimensions that still preserves second-order field equations, as in Ref.\cite{Lovelock gravity generic}.}
\begin{equation}\label{Einstein-Lovelock action}
\begin{split}
\mathcal{S}_{\text{L}}&=\int d^4x \sqrt{-g}\,\mathscr{L}_{\text{L}}+\mathcal{S}_m\quad\mbox{with}\\
\mathscr{L}_{\text{L}}&=\frac{1}{16\pi G}\left(
 R+\frac{a}{\sqrt{-g}} {}^*RR +b\mathcal{G}\right) ,
\end{split}
\end{equation}
where for simplicity we switch to the denotation
\begin{equation}\label{Redefinition CP}
{}^*RR \coloneqq\frac{1}{2} \epsilon_{\alpha\beta\mu\nu}
R^{\mu\nu}_{\;\;\;\;\,\gamma\delta}R^{\alpha\beta\gamma\delta}
\end{equation}
for the Chern-Pontryagin density, as the symbol ${}^*RR $ has been widely used in the literature of Chern-Simons gravity \cite{Chern-Simons 1, Chern-Simons 1.6, Chern-Simons 2}. In Eqs.(\ref{Lanczos-Lovelock action}), (\ref{Preparations Hilbert action}) and (\ref{Einstein-Lovelock action}), the matter action $\mathcal{S}_m$ is given in terms of the matter Lagrangian density $\mathscr{L}_m$ by $\mathcal{S}_m =\int d^4x \sqrt{-g}\,\mathscr{L}_m$,
and the stress-energy-momentum density tensor $T^{\text{(m)}}_{\mu\nu}$ is defined in the usual way by \cite{MTW Gravitation}
\begin{equation}\label{continuity conservation I}
\begin{split}
&\delta \mathcal{S}_m   =-\frac{1}{2}\int  d^4x \sqrt{-g}\,T_{\mu\nu}^{\text{(m)}} \delta g^{\mu\nu}\;\;\text{ with }\\
&T_{\mu\nu}^{\text{(m)}}\,\coloneqq\, \frac{-2}{\sqrt{-g}} \,\frac{\delta\,
\Big(\!\!\sqrt{-g}\,\mathscr{L}_m \Big)}{\delta g^{\mu\nu}}.
\end{split}
\end{equation}

The indistinguishability between $\mathcal{S}_{\text{L}}$ and
$\mathcal{S}_{\text{HE}}$ from their field equations begs the question: Does Einstein's equation come from $\mathcal{S}_{\text{L}}$ or $\mathcal{S}_{\text{HE}}$? Is there any way to
discriminate them?

Recall that GR from $\mathcal{S}_{\text{HE}}$ has a fundamental scalar-tensor counterpart, the Brans-Dicke gravity  \cite{Brans Dicke},
\begin{equation}\label{Standard Brans-Dicke action}
\begin{split}
\mathcal{S}_{\text{BD}}&=\int d^4x\sqrt{-g}\,\mathscr{L}_{\text{BD}}+ \mathcal{S}_m\quad\mbox{with}\\
\mathscr{L}_{\text{BD}}&=\frac{1}{16\pi}\left(\phiup R
-\frac{\omega_{\text{BD}}}{\phiup}\,\nabla_\alpha \phiup \nabla^\alpha\phiup \right)\,,
\end{split}
\end{equation}
which proves to be a successful alternative to GR that passes all typical GR tests   \cite{Equivalence principle}, and it is related to GR by
\begin{equation}
\begin{split}
\mathscr{L}_{\text{HE}}&=  \frac{R}{16\pi G}  \\
\Rightarrow \:\mathscr{L}_{\text{BD}}&= \frac{1}{16\pi}\,\left(\phiup R
-\frac{\omega_{\text{BD}}}{\phiup}\,\nabla_\alpha \phiup \nabla^\alpha\phiup\right) \,.
\end{split}
\end{equation}
That is to say,
Brans-Dicke firstly replaces the matter-gravity
coupling constant $G$ with a pointwise  scalar field $\phiup(x^\alpha)$ in accordance with the spirit of Mach's principle, $G\mapsto \phiup^{-1}$, and further adds to the action a formally canonical kinetic  term
$\displaystyle -\frac{\omega_{\text{BD}}}{\phiup}\,\nabla_\alpha \phiup \nabla^\alpha\phiup$ governing the kinetics of $\phiup(x^\alpha)$.
Applying this prescription to the Lovelock action Eq.(\ref{Einstein-Lovelock action}), we obtain
\begin{eqnarray}\label{LBD action 0}
\mathscr{L}_{\text{L}}= \frac{1}{16\pi G}&&\left( R
+\frac{a}{\sqrt{-g}} {}^*RR + b\mathcal{G}\right) \\
\Rightarrow \mathscr{L}_{\text{LBD}}=  \frac{1}{16\pi}\Bigg[ \phiup\, &&\left(R
+\frac{a}{\sqrt{-g}} {}^*RR + b\mathcal{G}\right) -\frac{\omega_{\text{L}}}{\phiup}\nabla_\alpha \phiup \nabla^\alpha\phiup \Bigg],\nonumber
\end{eqnarray}
where the Lovelock parameter $\omega_{\text{L}}$ is a dimensionless constant. Based on Eq.(\ref{LBD action 0}),
we obtain what we dub as the \emph{Lovelock-Brans-Dicke} (henceforth LBD) gravity with the action
\begin{equation}\label{LBD action}
\mathcal{S}_{\text{LBD}}=\int d^4x\sqrt{-g}\,\mathscr{L}_{\text{LBD}}+ \mathcal{S}_m\,,
\end{equation}
or the \emph{Lanczos-Lovelock-Brans-Dicke gravity}, as Lovelock's theorem is based on Lanczos' discovery that an isolated ${}^*RR$ or $\mathcal{G}$ in the action does not affect the field equation \cite{Lanczos original paper}.

Unlike the
$ \delta({}^*RR )/\delta g^{\mu\nu}$ and $\delta (\!\sqrt{-g}\mathcal{G})/\delta g^{\mu\nu} $ in $\delta\mathcal{S}_{\text{L}}/\delta g^{\mu\nu}$,  the
$\left[\phiup \delta({}^*RR )\right]/\delta g^{\mu\nu}$ and $\left[\phiup \delta (\!\sqrt{-g}\mathcal{G}) \right]/\delta g^{\mu\nu} $
for  $\delta\mathcal{S}_{\text{LBD}}/\delta g^{\mu\nu}$
are no longer pure divergences, because the scalar field  $\phiup(x^\alpha)$ as a nontrivial coefficient will be absorbed into the variations of ${}^*RR $ and $\sqrt{-g}\mathcal{G}$ when integrating by parts. Hence, although $\mathcal{S}_{\text{L}}$ and $\mathcal{S}_{\text{HE}}$ are indistinguishable,
their respective scalar-tensor counterparts $\mathcal{S}_{\text{LBD}}$ and $\mathcal{S}_{\text{BD}}$ are different.

Note that the cosmological-constant term $-2\Lambda$ in Eq.(\ref{Lanczos-Lovelock action}) is temporarily abandoned in $\mathscr{L}_{\text{L}}$; otherwise, it would add an extra term $-2\Lambda\phiup$
to $\mathscr{L}_{\text{LBD}}$, which serves as a simplest linear potential.  This is primarily for a better analogy between the LBD and the Brans-Dicke gravities, as the latter in its standard form does not contain a potential term $V(\phiup)$, and an unspecified potential $V(\phiup)$ would cause too much arbitrariness to $\mathscr{L}_{\text{LBD}}$.

Also, Lovelock's original action Eq.(\ref{Lanczos-Lovelock action}) concentrates on the \textit{algebraic} curvature invariants; in fact, one can further add to
Eq.(\ref{Lanczos-Lovelock action}) the relevant differential  terms $\Box R$, $\Box {}^*RR$,  and $\Box \mathcal{G}$ ($\Box=g^{\alpha\beta}\nabla_\alpha\nabla_\beta$ denoting the covariant d'Alembertian), while the field equation will remain unchanged. This way, the gravitational Lagrangian density in Eq.(\ref{Einstein-Lovelock action}) is enriched into
\begin{equation}\label{Modified LBD action I}
\begin{split}
\mathscr{L}=\frac{1}{16\pi G}\left(
 R+\frac{a{}^*RR}{\sqrt{-g}} +b\mathcal{G} +c\Box R+d\Box {}^*RR+e \Box \mathcal{G}\right) ,
\end{split}
\end{equation}
with $\{c,d,e\}$ being constants, and its Brans-Dicke-type counterpart extends Eq.(\ref{LBD action 0}) into
\begin{eqnarray} \label{Modified LBD action II}
\begin{split}
\mathscr{L}=  \frac{1}{16\pi}\Bigg[ &\phiup\,  \left(R
+\frac{a}{\sqrt{-g}} {}^*RR + b\mathcal{G}\right)
 -\frac{\omega_{\text{L}}}{\phiup}\nabla_\alpha \phiup \nabla^\alpha\phiup \\
& +\phiup\,  \bigg(c\Box R+d\Box {}^*RR+e \Box \mathcal{G}   \bigg)\Bigg]\,,
\end{split}
\end{eqnarray}
where $\phiup \cdot\left(c\Box R+d\Box {}^*RR+e \Box \mathcal{G}   \right)$ have nontrivial contributions to the field equation. However, unlike $*RR$ and $\sqrt{-g}\mathcal{G}$ which are divergences of their respective topological current as in Eqs.(\ref{CP current}) and (\ref{GB current}), \{$\Box R$, $\Box {}^*RR$, $\Box \mathcal{G}$\} are total derivatives simply because the d'Alembertian $\Box$ satisfies $\sqrt{-g}\Box \Theta=
\partial_\alpha\left(\sqrt{-g}\,g^{\alpha\beta}\partial_\beta \Theta\right)$ when acting on an arbitrary scalar field $\Theta$; in this sense, these differential boundary terms which contain fourth-order derivatives of the metric are less interesting than  $*RR$ and $\mathcal{G}$. In this paper, we will focus on the LBD gravity $\mathscr{L}_{\text{LBD}}$ Eq.(\ref{LBD action 0}) built upon the original Lovelock action and Lovelock's theorem, rather than Eq.(\ref{Modified LBD action II}) out of the modified action Eq.(\ref{Modified LBD action I}).


\section{Gravitational and wave equations}\label{Gravitational and wave equations LBD}

In this section we will work out the gravitational field equation $\delta\mathcal{S}_{\text{LBD}}/\delta g^{\mu\nu}=0$ and the wave equation $\delta\mathcal{S}_{\text{LBD}}/\delta\phiup=0$ for the LBD gravity. First of all, with $\delta g_{\alpha\beta}= -g_{\alpha\mu} g_{\beta\nu}\delta g^{\mu\nu}$,  $\delta \Gamma^\lambda_{\alpha\beta}=\frac{1}{2} g^{\lambda\sigma}\left(\nabla_\alpha \delta g_{\sigma\beta}+\nabla_\beta \delta g_{\sigma\alpha}-\nabla_\sigma \delta g_{\alpha\beta} \right)$, and the Palatini identity $\delta R^\lambda_{\;\;\alpha\beta\gamma}=\nabla_\beta \left(\delta \Gamma^\lambda_{\gamma\alpha}\right)-\nabla_\gamma \left(\delta\Gamma^\lambda_{\beta\alpha}\right)$ \cite{DeWitt Specific Lagrangians},
for the first term $\phiup R$ in $\mathscr{L}_{\text{LBD}}$ it is easy to work out that
\begin{equation}\label{vary phi R}
\frac{1}{\sqrt{-g}}\frac{\delta (\!\sqrt{-g}\,\phiup R)}{\delta g^{\mu\nu}} \cong -\frac{1}{2}\phiup Rg_{\mu\nu}+ \phiup R_{\mu\nu}+
\left(g_{\mu\nu} \Box-\nabla_\mu\nabla_\nu\right) \phiup ,
\end{equation}
where
$\cong$ means equality by neglecting all total-derivative terms which are boundary terms for the action.


\subsection{Coupling to the Chern-Pontryagin invariant}\label{Sec Coupling to the Chern-Pontryagin invariant}

The Chern-Pontryagin density ${}^*RR$ in $\mathscr{L}_{\text{LBD}}$ measures the gravitational effects of parity violation through $\int  d^4x\, \phiup {}^*RR$ for its dependence on the Levi-Civita pseudotensor. In addition to Eq.(\ref{Redefinition CP}),
${}^*RR$ is related to the left dual of the Riemann tensor via
\begin{equation}\label{Dual CP}
{}^*RR
= \frac{1}{2}\left(\epsilon_{\alpha\beta\mu\nu} R^{\mu\nu}_{\;\;\;\;\gamma\delta}\right) R^{\alpha\beta\gamma\delta}
=  {}^{*}R_{\alpha\beta\gamma\delta} R^{\alpha\beta\gamma\delta}\,.
\end{equation}
Applying the Ricci decomposition $R_{\alpha\beta\gamma\delta}=C_{\alpha\beta\gamma\delta}+\frac{1}{2}\left( g_{\alpha\gamma}R_{\beta\delta}-\right.$ $\left.g_{\alpha\delta}R_{\beta\gamma}+
g_{\beta\delta}R_{\alpha\gamma}-g_{\beta\gamma}R_{\alpha\delta} \right)
-\frac{1}{6}\left( g_{\alpha\gamma} g_{\beta\delta}
- g_{\alpha\delta} g_{\beta\gamma}  \right)R$ to Eq.(\ref{Dual CP}) and using the cyclic identity $C_{\alpha\beta\gamma\delta}+C_{\alpha\gamma\delta\beta}+C_{\alpha\delta\beta\gamma}=0$ for the traceless Weyl tensor, one could find the equivalence
\begin{equation}\label{CP RC equivalence}
{}^*RR={}^*CC
 \coloneqq\frac{1}{2}\left(\epsilon_{\alpha\beta\mu\nu}
 C^{\mu\nu}_{\;\;\;\;\gamma\delta}\right) C^{\alpha\beta\gamma\delta}
=  {}^{*}C_{\alpha\beta\gamma\delta} C^{\alpha\beta\gamma\delta}\,,
\end{equation}
which indicates that the Chern-Pontryagin density is conformally invariant \cite{Chern Simons scalar} under a rescaling $g_{\mu\nu}\mapsto \Omega(x^\alpha)^2 \cdot g_{\mu\nu}$ of the metric tensor.

With the Chern-Simons topological current $K^\mu$ in Eq.(\ref{CP current}), one can integrate by parts and obtain
$\int d^4x\, \phiup {}^*RR =-4\int d^4x\,\phiup \left(\partial_\mu K^\mu  \right)
=-4\int d^4x\, \partial_\mu \left(\phiup K^\mu  \right)
+4\int d^4x\, \left(\partial_\mu \phiup \right) K^\mu $.
Hence, instead of directly varying $\phiup {}^*RR$ with respect to the inverse metric, we firstly vary the four-current $K^\mu$ by the Levi-Civita connection. It follows that
\begin{eqnarray}\label{Vary CP Jackiw}
& &\delta\int d^4x\,\phiup {}^*RR \;\cong\; 4\int d^4x\,\left(\partial_\mu \phiup \right) \delta K^\mu  \nonumber\\
= &&2\int  d^4x\,\left(\partial_\mu \phiup \right)\epsilon^{\mu\alpha\beta\gamma}R^\xi_{\;\;\,\rho\beta\gamma}\delta \Gamma^\rho_{\alpha\xi}\nonumber\\
= & &2\int d^4x\, \left(\partial_\mu \phiup \right)\epsilon^{\mu\alpha\beta\gamma}R^{\xi\nu}_{\;\;\;\;\beta\gamma}\left(\nabla_\xi \delta g_{\alpha\nu}-\nabla_\nu \delta g_{\alpha\xi}\right)\nonumber\\
\cong&-&2\int d^4x\,\left[\left(\partial_\mu \phiup \right)\epsilon^{\mu\alpha\beta\gamma}\nabla_\xi R^{\xi\nu}_{\;\;\;\;\beta\gamma} + \left(\partial_\mu \partial_\xi  \phiup \right) \epsilon^{\mu\alpha\beta\gamma}R^{\xi\nu}_{\;\;\;\;\beta\gamma}   \right]\delta g_{\alpha\nu} \nonumber\\
=&-&4\int d^4x\,\left[\left(\partial_\mu \phiup \right)\epsilon^{\mu\alpha\beta\gamma}\nabla_\beta R^{\;\;\nu}_{\gamma} + \left(\partial_\mu \partial_\xi  \phiup \right) {}^*R^{\mu\alpha\xi\nu}   \right] \delta g_{\alpha\nu} \\
=&&4\int d^4x\,\left[\left(\partial^\mu \phiup \right)\epsilon_{\mu\alpha\beta\gamma}\nabla^\beta R^{\gamma}_{\;\;\nu} + \left(\partial_\mu \partial_\xi  \phiup \right) {}^*R^{\mu\;\;\,\xi}_{\;\;\,\alpha\;\;\nu}   \right] \delta g^{\alpha\nu} ,\label{Vary CP last step}
\end{eqnarray}
where, in the third row we expanded $\delta \Gamma^\rho_{\alpha\xi}$ and made use of the cancelation $R^{\xi\nu}_{\;\;\;\;\beta\gamma}\nabla_\alpha \delta g_{\xi\nu}=0$ due to the skew-symmetry for  the indices $\xi\leftrightarrow\nu$; in the fourth row, we
applied the replacement $\nabla_\xi R^{\xi\nu}_{\;\;\;\;\beta\gamma}=\nabla_\beta R^{\;\;\nu}_{\gamma}-\nabla_\gamma R^{\;\;\nu}_{\beta}$ in accordance with the relation
\begin{equation}\label{Bianchi implications 0}
\nabla^\alpha R_{\alpha\mu\beta\nu}=\nabla_\beta R_{\mu\nu}-\nabla_\nu R_{\mu\beta},
\end{equation}
which is an implication of the second Bianchi identity $\nabla_\gamma R_{\alpha\mu\beta\nu}$ $+\nabla_\nu R_{\alpha\mu\gamma\beta}+\nabla_\beta R_{\alpha\mu\nu\gamma}=0$; in the last step, we raised the indices of $\delta g_{\alpha\nu} $ to $\delta g^{\alpha\nu}  $ and thus had the overall minus sign dropped. In Eq.(\ref{Vary CP last step}) we adopted the usual notation $\partial^\mu \phiup \equiv g^{\hat{\mu}\mu}\partial_{\hat{\mu}} \phiup$, and note that  $\left(\partial_\mu \partial_\xi  \phiup \right) {}^*R^{\mu\;\;\,\xi}_{\;\;\,\alpha\;\;\nu} \neq   \left(\partial^\mu \partial^\xi  \phiup\right)  {}^*R_{\mu\alpha\xi\nu} $ since in general the metric tensor does not commute with partial derivatives and thus $\partial^\mu \partial^\xi  \phiup=g^{\mu\hat{\mu}} \partial_{\hat{\mu}}\left(g^{\xi \hat{\xi}} \partial_{\hat{\xi}}  \phiup  \right)
\neq g^{\mu\hat{\mu}} g^{\xi \hat{\xi}} \partial_{\hat{\mu}} \partial_{\hat{\xi}} \phiup $.
Relabel the indices of Eq.(\ref{Vary CP Jackiw}) and we obtain the variational derivative
\begin{equation*}\label{Vary CP defH}
\frac{1}{\sqrt{-g}}\frac{\delta \left(\phiup {}^*RR\right)}{\delta g^{\mu\nu}}\,\eqqcolon\, H_{\mu\nu}^{\text{(CP)}}
\;\;\;\mbox{and}
\end{equation*}
\begin{equation}\label{Vary CP}
\begin{split}
\sqrt{-g}\,H_{\mu\nu}^{\text{(CP)}}= &2\partial^\xi\phiup\cdot
\left( \epsilon_{\xi\mu\alpha\beta}\nabla^\alpha R^\beta_{\;\;\nu} +
\epsilon_{\xi\nu\alpha\beta}\nabla^\alpha R^\beta_{\;\;\mu} \right) \\
&+2\partial_\alpha \partial_\beta  \phiup \cdot \left({}^*R^{\alpha\;\;\,\beta}_{\;\;\,\mu\;\;\nu}
+{}^*R^{\alpha\;\;\,\beta}_{\;\;\,\nu\;\;\mu}\right).
\end{split}
\end{equation}
Compared with Eq.(\ref{vary phi R}), $H_{\mu\nu}^{\text{(CP)}}$ does not contain a $-\frac{1}{2}\phiup {}^*RR g_{\mu\nu}$ term, because ${}^*RR$ by itself already serves as a covariant density as opposed to the usual form $\sqrt{-g}\, \mathcal{R}$ for other curvature invariants.

Note that the nonminimal coupling between a scalar field and ${}^*RR$ is crucial to the Chern-Simons gravity; however, its original proposal Ref. \cite{Chern-Simons 1} had adopted the opposite geometric system which uses the metric signature $(+,---)$, the conventions \{$R^{\alpha}_{\;\;\,\beta\gamma\delta}=\partial_\delta \Gamma^\alpha_{\;\;\gamma\beta}\cdots$, $R_{\mu\nu}=R^\alpha_{\;\;\mu\alpha\nu}$\}, Einstein's equation $R_{\mu\nu}-\frac{1}{2}Rg_{\mu\nu}=-8\pi GT_{\mu\nu}^{\text{(m)}}$, and the definition ${}^*RR= - {}^{*}R_{\alpha\beta\gamma\delta}\,R^{\alpha\beta\gamma\delta}
=-\frac{1}{2}\Big(\epsilon_{\alpha\beta\mu\nu} R^{\mu\nu}_{\;\;\;\;\gamma\delta}\Big)\,R^{\alpha\beta\gamma\delta}$.  This has caused quite a few mistakes in the subsequent Chern-Simons literature that adopt different conventions, and we hope the details in this subsection could correct these misunderstandings. Also,
in Eq. (\ref{Vary CP}), the quantities
$\{\epsilon_{\xi\mu\alpha\beta}$, $K^\mu$, ${}^*RR$, $R^\beta_{\;\;\nu}\,,{}^*R_{\beta\mu\alpha\nu} \}$ have the same values in both sets of sign conventions.
See our note Ref.\cite{Convention switch} for further clarification of this issue.


\subsection{Coupling to the Gauss-Bonnet invariant}\label{Sec Coupling to the Gauss-Bonnet invariant}

The third term $\phiup \mathcal G$ in $\mathscr{L}_{\text{LBD}}$ represents the nonminimal coupling between the scalar field and the Gauss-Bonnet invariant $\mathcal G= R^2-4R_c^2+R_m^2$,
where we have employed the straightforward abbreviations $R_c^2\coloneqq R_{\alpha\beta}R^{\alpha\beta}$ and $R_m^2\coloneqq R_{\alpha\mu\beta\nu}R^{\alpha\mu\beta\nu}$ to  denote the Ricci and Riemann tensor squares. Following the standard procedures of  variational derivative as before in $\delta\left(\sqrt{-g}\phiup R\right)/\delta g^{\mu\nu}$, we have
\begin{equation}\label{GB terms}
\frac{\delta\left(\!\sqrt{-g}\phiup\mathcal{G}\right)}{\sqrt{-g}\;\delta g^{\mu\nu}}=
\frac{\delta(\phiup R^2)}{\delta g^{\mu\nu}}-4 \frac{\delta(\phiup R_c^2)}{\delta g^{\mu\nu}}
+ \frac{\delta(\phiup R_m^2)}{\delta g^{\mu\nu}}-\frac{1}{2}\phiup\mathcal{G} g_{\mu\nu},
\end{equation}
with
\begin{eqnarray}
\frac{\delta\left(\phiup R^2\right)}{\delta g^{\mu\nu}}\cong&& \,2\phiup RR_{\mu\nu}
+2 \left(g_{\mu\nu}\Box-\nabla_\mu\!\nabla_\nu\right)\,\left(\phiup R\right) \label{GB terms I}\\
\frac{\delta\left(\phiup R_c^2\right)}{\delta g^{\mu\nu}} \cong&& \,2\phiup R_\mu^{\;\;\,\alpha}R_{\alpha\nu}
+\Box \left(\phiup R_{\mu\nu}\right)- \nabla_\alpha\!\nabla_{\nu} \left(\phiup R_{\mu}^{\;\;\alpha}\right) \nonumber \\
&&-\nabla_\alpha\!\nabla_{\mu} \left(\phiup R_{\nu}^{\;\;\alpha}\right) + g_{\mu\nu} \nabla_\alpha\!\nabla_\beta \left(\phiup  R^{\alpha\beta}  \right)\label{GB terms II}\\
\frac{\delta\left(\phiup R_m^2\right)}{\delta g^{\mu\nu}} \cong&& \,2\phiup R_{\mu\alpha\beta\gamma}R_{\nu}^{\;\;\,\alpha\beta\gamma}
+4 \nabla^\beta \nabla^\alpha \left(\phiup  R_{\alpha\mu \beta\nu} \right),\label{GB terms III}
\end{eqnarray}
where total-derivative terms have been removed.
Recall that besides  Eq.(\ref{Bianchi implications 0}), the second Bianchi identity also has the following implications
which transform the derivative of a high-rank curvature tensor into that of lower-rank tensors plus nonlinear algebraic terms:
\begin{eqnarray}
\nabla^\alpha R_{\alpha\beta} &=& \frac{1}{2}\,\nabla_\beta R \label{Bianchi implications}   \\
\nabla^\beta\nabla^\alpha  R_{\alpha\beta} &=& \frac{1}{2}\,\Box R  \\
\nabla^\beta \nabla^\alpha  R_{\alpha\mu\beta\nu}\,=\,\Box R_{\mu\nu} &-& \frac{1}{2}\nabla_\mu\! \nabla_\nu
R+R_{\alpha\mu\beta\nu}R^{\alpha\beta}-R_{\mu}^{\;\;\,\alpha}R_{\alpha\nu} \\
\hspace{-1mm}\nabla^\alpha\nabla_\mu R_{\alpha\nu}+\nabla^\alpha\nabla_\nu R_{\alpha\mu} &=& \nabla_\mu\!\nabla_\nu R
-2R_{\alpha\mu\beta\nu}R^{\alpha\beta}+2R_{\mu}^{\;\;\,\alpha}R_{\alpha\nu}.\label{Bianchi implications end}
\end{eqnarray}
Using Eq.(\ref{Bianchi implications 0}) and Eqs.(\ref{Bianchi implications})-(\ref{Bianchi implications end}) to expand the second-order covariant derivatives in Eqs.(\ref{GB terms I})-(\ref{GB terms III}), and putting them back into Eq.(\ref{GB terms}), we obtain
\begin{equation*}
\frac{1}{\sqrt{-g}}\frac{\delta(\sqrt{-g}\,\phiup\,\mathcal{G})}{\delta
g^{\mu\nu}}\,\eqqcolon\,H_{\mu\nu}^{\text{(GB)}}
\quad\mbox{with}
\end{equation*}
\begin{eqnarray}
H_{\mu\nu}^{\text{(GB)}}
=&&\phiup\left( 2R R_{\mu\nu}-4  R_\mu^{\;\;\,\alpha}R_{\alpha\nu}-4 R_{\alpha\mu\beta\nu}R^{\alpha\beta} +2
R_{\mu\alpha\beta\gamma}R_{\nu}^{\;\;\,\alpha\beta\gamma}\right)\nonumber\\
&&+2R\left(g_{\mu\nu}\Box
-\nabla_\mu\!\nabla_\nu\right) \phiup
+4R_{\mu}^{\;\;\,\alpha}\nabla_\alpha\!\nabla_{\nu}\phiup +4R_{\nu}^{\;\;\,\alpha}\nabla_\alpha\!\nabla_{\mu} \phiup\nonumber\\
&&-4R_{\mu\nu}\Box \phiup -4g_{\mu\nu}  R^{\alpha\beta}\nabla_\alpha\!\nabla_\beta \phiup+4 R_{\alpha\mu \beta\nu} \nabla^\beta \nabla^\alpha  \phiup \nonumber\\
&&-\frac{1}{2} \phiup \mathcal{G} g_{\mu\nu},\label{GB variation II}
\end{eqnarray}
where the second-order derivatives $\{\Box,\nabla_\alpha\!\nabla_\nu, \text{etc}\}$  only act on the scalar field $\phiup$.

However, we realize that Eq.(\ref{GB variation II}) is still not the ultimate expression.
In four dimensions, $\sqrt{-g}\,\mathcal{G}$ is proportional to the Euler-Poincar\'e topological density,
$\mathcal{G}=\left(\frac{1}{2}\epsilon_{\alpha\beta\gamma\zeta}R^{\gamma\zeta\eta\xi}\right)\cdot \left(\frac{1}{2}\epsilon_{\eta\xi\rho\sigma} R^{\rho\sigma\alpha\beta}\right)
={}^*R_{\alpha\beta}^{\;\;\;\;\,\eta\xi}\cdot{}^*R_{\eta\xi}^{\;\;\;\;\,\alpha\beta}$,
and the integral $ \frac{1}{32\pi^2}\int dx^4 \sqrt{-g}\,\mathcal{G}$ equates the Euler characteristic $\chiup(\mathcal{M})$ of the spacetime. Thus
$\frac{\delta}{\delta g^{\mu\nu}}\int dx^4  \sqrt{-g}\,\mathcal{G} = 32\pi^2\frac{\delta }{\delta g^{\mu\nu}}\chiup(\mathcal{M})  \equiv 0$.
Based on Eqs.(\ref{GB terms I})-(\ref{GB terms III}), one could easily obtain the Bach-Lanczos identity from the explicit variational derivative $\delta\left(\sqrt{-g}\,\mathcal{G}\right)/\delta g^{\mu\nu}$,
\begin{equation}\label{GaussBonnet Bach-Lanczos identity II}
2 RR_{\mu\nu}-4 R_\mu^{\;\;\,\alpha}R_{\alpha\nu}-4 R_{\alpha\mu\beta\nu}R^{\alpha\beta}
+2R_{\mu\alpha\beta\gamma}R_{\nu}^{\;\;\,\alpha\beta\gamma}\equiv \frac{1}{2}\mathcal{G}g_{\mu\nu},
\end{equation}
with which Eq.(\ref{GB variation II}) can be best simplified into
\begin{eqnarray}
H_{\mu\nu}^{\text{(GB)}}=
&&2R\left(g_{\mu\nu}\Box
-\nabla_\mu\!\nabla_\nu\right)\phiup +4R_{\mu}^{\;\;\,\alpha}\nabla_\alpha\!\nabla_{\nu}\phiup +4R_{\nu}^{\;\;\,\alpha}\nabla_\alpha\!\nabla_{\mu} \phiup \nonumber\\
&&-4R_{\mu\nu}\Box \phiup -4g_{\mu\nu} \cdot R^{\alpha\beta}\nabla_\alpha\!\nabla_\beta \phiup+4R_{\alpha\mu \beta\nu}
\nabla^\beta \nabla^\alpha  \phiup, \label{GB variation III}
\end{eqnarray}
whose trace is
\begin{equation} \label{GB variation trace}
g^{\mu\nu} H_{\mu\nu}^{\text{(GB)}}= 2R \Box \phiup - 4R^{\alpha \beta } \nabla_\alpha  \nabla_\beta \phiup.
\end{equation}

In the existent literature, the effects of the generalized and thus nontrivial Gauss-Bonnet dependence for the field equations are generally depicted in the form analogous to Eq.(\ref{GB variation II}), such as the string-inspired  Gauss-Bonnet effective dark energy \cite{GaussBonnet dark energy} with $\mathscr{L}=\frac{1}{16\pi G}R-\frac{\gamma}{2}\partial_\mu\varphi \partial^\mu\varphi-V(\varphi)+f(\varphi)\mathcal{G}$, as well as the $R+f(\mathcal{G})$ \cite{GaussBonnet first model R/2k+f(G)}, the $f(R,\mathcal{G})$ \cite{GaussBonnet second model f(R G)+Lm} and the $f(R,\mathcal{G}, \mathscr{L}_m)$ \cite{AA Tian-Booth Paper} generalized Gauss-Bonnet gravities. Here we emphasize that the Gauss-Bonnet effects therein could all be simplified into the form of Eq.(\ref{GB variation III}).


\subsection{Gravitational field equation}

Collecting the results in Eqs.(\ref{vary phi R}), (\ref{Vary CP}), and (\ref{GB variation III}),
we finally obtain the gravitational field equation
\begin{equation}\label{BDLL full 1}
\begin{split}
\phiup &\left(R_{\mu\nu} -\frac{1}{2}  R  g_{\mu\nu}\right)
-\frac{\omega_{\text{L}}}{\phiup}\left(\nabla_\mu \phiup \nabla_\nu \phiup-\frac{1}{2}g_{\mu\nu}\nabla_\alpha \phiup \nabla^\alpha\phiup\right)\\
+ &\left(g_{\mu\nu}\Box -\nabla_\mu \nabla_\nu \right) \phiup  + a  H_{\mu\nu}^{\text{(CP)}} +b H_{\mu\nu}^{\text{(GB)}} =8\pi  T_{\mu\nu}^{\text{(m)}},
\end{split}
\end{equation}
where $ H_{\mu\nu}^{\text{(CP)}}$ vanishes for all spherically symmetric or conformal flat spacetimes. 
Eq.(\ref{BDLL full 1}) yields the trace equation
\begin{equation}\label{BDLL full 2 trace}
-\phiup R
+\frac{\omega_{\text{L}}}{\phiup}\nabla_\alpha \phiup \nabla^\alpha\phiup
+\left(3+2bR\right) \Box \phiup   -4 b R^{\alpha\beta}\nabla_\alpha\!\nabla_\beta \phiup =8\pi  T^{\text{(m)}},
\end{equation}
where $H_{\mu\nu}^{\text{(CP)}}$ is always traceless, $g^{\mu\nu}H_{\mu\nu}^{\text{(CP)}}\equiv0$ -- this is not a surprise because it equivalently traces back to the effects of the dual square ${}^*CC$ of the traceless Weyl tensor.

Note that in existent studies the invariants ${}^*RR$ and $\mathcal{G}$ have demonstrated their importance in various aspects. For example, as shown by Eq.(6) of Ref.\cite{Weinberg Effective Field} [recall the equivalence ${}^*RR={}^*CC$ in Eq.(\ref{CP RC equivalence})], in the effective field theory for the initial cosmic inflation, the only leading-order fluctuations to the standard inflation action in the tensor modes are the parity-violation Chern-Pontryagin and the topological Gauss-Bonnet effects.


\subsection{Wave equations}

Straightforward extremization of $\mathcal{S}_{\text{LBD}}$ with respect to the scalar field
yields the \emph{kinematical} wave equation
\begin{equation}\label{LBD wave kinematic}
\frac{2\omega_{\text{L}} }{\phiup} \Box\phiup  = -R+\frac{ \omega_{\text{L}}}{\phiup^2}
\nabla_{\alpha}\phiup \nabla^{\alpha} \phiup -\left(\frac{a}{\sqrt{-g}} {}^*RR +b \mathcal{G}\right),
\end{equation}
with $\Box\phiup=\frac{1}{\sqrt{-g}}
\partial_\alpha\left(\sqrt{-g}\,g^{\alpha\beta}\partial_\beta \phiup\right)$. We regard Eq.(\ref{LBD wave kinematic}) as ``kinematical'' because  it does not explicitly relate the propagation of $\phiup$ to the matter distribution $\mathscr{L}_m$ or $T^{\text{(m)}}=g^{\mu\nu}T^{\text{(m)}}_{\mu\nu}$.

Combine Eq.(\ref{LBD wave kinematic}) with the gravitational trace equation (\ref{BDLL full 2 trace}), and it follows that
\begin{equation}\label{LBD wave dynamical}
\begin{split}
\left(2\omega_{\text{L}} +3+2bR\right)\Box\phiup = &-\left(
\frac{a}{\sqrt{-g}} {}^*RR +b \mathcal{G}\right)\phiup\\
 &+ 8\pi T^{\text{(m)}}+ 4 b R^{\alpha\beta}\nabla_\alpha\!\nabla_\beta \phiup,
\end{split}
\end{equation}
which serves as the generalized Klein-Gordon equation that governs the dynamics of the scalar field.


\section{The $\omega_{\text{L}}\to\infty$ limit and GR}\label{Sec The omega to infty limit and GR}
From the dynamical equation (\ref{LBD wave dynamical}), we obtain
\begin{eqnarray}
 \Box\phiup
= \frac{1}{2\omega_{\text{L}} +3+2bR}\Bigg\{-\left(
\frac{a}{\sqrt{-g}} {}^*RR +b \mathcal{G}\right)\phiup&&\nonumber \\
+ 8\pi  T^{\text{(m)}}+4 b R^{\alpha\beta}\nabla_\alpha\!\nabla_\beta \phiup \Bigg\}\label{LBD wave omega limit}&&.
\end{eqnarray}
The topology-gravity coupling strengths $\{a,b\}$ should take finite values -- just like the Newtonian constant $G$ for matter-gravity coupling. Similarly the curvature invariants
$\{R, {}^*RR, \mathcal{G} \}$ for a physical spacetime should be finite, and we further assume the scalar field $\phiup$ to be nonsingular. Thus, in the limit $\omega_{\text{L}}\to\infty$, Eq.(\ref{LBD wave omega limit}) yields $\Box\phiup= \mathds{O}\left(\frac{1}{\omega_{\text{L}}}\right)$ and
\begin{equation}\label{phiup omega infinity limit}
\phiup=\langle\phiup\rangle+\mathds{O}\left(\frac{1}{\omega_{\text{L}}}\right)=\frac{1}{G}+\mathds{O}\left(\frac{1}{\omega_{\text{L}}}\right),
\end{equation}
where $\langle\phiup\rangle$ denotes the expectation value of the scalar field and we expect it to be the inverse of the Newtonian constant $1/G$. Under the behaviors Eq.(\ref{phiup omega infinity limit}) in the infinite $\omega_{\text{L}}$ limit, we have $H_{\mu\nu}^{\text{(CP)}}=0 =H_{\mu\nu}^{\text{(GB)}}$, and the field equation (\ref{BDLL full 1}) reduces to become Einstein's equation $R_{\mu\nu} -\frac{1}{2}  R  g_{\mu\nu} =8\pi G T_{\mu\nu}^{\text{(m)}}$.

On the other hand, from Eq.(\ref{LBD wave omega limit}) we can also observe that $\Box\phiup\equiv 0$ in the special situation
\begin{equation} \label{topological balance condition}
-4 b  R^{\alpha\beta}\nabla_\alpha\!\nabla_\beta \phiup +\left(
\frac{a}{\sqrt{-g}}{}^*RR +b \mathcal{G}\right)\phiup=8\pi  T^{\text{(m)}},
\end{equation}
and the scalar field becomes undeterminable from the dynamical equation (\ref{LBD wave omega limit}).

The term $-4 b  R^{\alpha\beta}\nabla_\alpha\!\nabla_\beta \phiup$ comes from the trace $g^{\mu\nu} H_{\mu\nu}^{\text{(GB)}}$, while  ${}^*RR$ and $\mathcal{G}$ are respectively related to the topological instanton number \cite{Chern Simons scalar} and the Euler characteristic. Thus, all terms on the left hand side of Eq.(\ref{topological balance condition}) are related to topological effects nonminimally coupled with $\phiup$, and they cancel out the trace of the matter tensor. In this sense, we call Eq.(\ref{topological balance condition}) the \emph{topological balance condition}.

Putting $\Box\phiup\equiv 0$ and the condition Eq.(\ref{topological balance condition}) back into the trace equation (\ref{BDLL full 2 trace}), we obtain
\begin{eqnarray} \label{topological balance condition kinematical}
\frac{ \omega_{\text{L}}}{\phiup^2}
\nabla_{\alpha}\phiup \nabla^{\alpha} \phiup   &=& R+ \frac{a}{\sqrt{-g}} {}^*RR +b \mathcal{G}\\
&=&-\omega_{\text{L}}\Box\ln \phiup,
\end{eqnarray}
where in the second step we further made use of the expansion $\Box\ln \phiup=\nabla^\alpha\left(\frac{1}{\phiup}\nabla_\alpha \phiup\right)=-\frac{1}{\phiup^2}\nabla_{\alpha}\phiup \nabla^{\alpha} \phiup+\frac{1}{\phiup}\Box\phiup=-\frac{1}{\phiup^2}\nabla_{\alpha}\phiup \nabla^{\alpha} \phiup$ for $\Box\phiup\equiv 0$.
Thus it follows that
\begin{equation}\label{phiup omega infinity limit Topo balanced}
\omega_{\text{L}}\nabla_{\alpha}\left(\ln\phiup\right) \nabla^{\alpha} \left(\ln\phiup\right)
= R+ \frac{a}{\sqrt{-g}} {}^*RR +b \mathcal{G}\,.
\end{equation}
For $\omega_{\text{L}}\to\infty$, this equation gives the estimate
\begin{equation}
\left\|\,\nabla_{\alpha}(\ln\phiup)\,\right\| \sim \sqrt{\frac{R+ \frac{a}{\sqrt{-g}} {}^*RR +b \mathcal{G}}{\omega_{\text{L}}}} \sim
\mathds{O}\left(\frac{1}{\sqrt{\omega_{\text{L}}}}\right),
\end{equation}
which integrates to yield $\ln\phiup = \mbox{constant}+
\mathds{O}\left(\frac{1}{\sqrt{\omega_{\text{L}}}}\right)$. Hence, $\phiup$ satisfies
\begin{equation}\label{Estimate infinite omega II}
\phiup \sim \phiup_0 +\mathds{O}\left(\frac{1}{\sqrt{\omega_{\text{L}}}}\right),
\end{equation}
where the constant $\phiup_0$ is the average value of $\phiup$. In accordance with Eq.(\ref{topological balance condition kinematical}) and the estimate Eq.(\ref{Estimate infinite omega II}), the term $-\frac{\omega_{\text{L}}}{\phiup}\left(\nabla_\mu \phiup \nabla_\nu \phiup-\frac{1}{2}g_{\mu\nu}\nabla_\alpha \phiup \nabla^\alpha\phiup\right)$ in the field equation (\ref{BDLL full 1}), which arises from the source $-\frac{\omega_{\text{L}}}{\phiup}\nabla_\alpha \phiup \nabla^\alpha\phiup$ in $\mathcal{S}_{\text{LBD}}$, will not vanish. This way, the $\omega_{\text{L}}\to\infty$ limit could not recover Einstein's equation and GR in situations where the topological balance condition Eq.(\ref{topological balance condition}) holds, although the existence of such solutions remains to be carefully checked.

This is similar to the Brans-Dicke theory given by the action Eq.(\ref{Standard Brans-Dicke action}), which recovers GR in  the limit $\omega_{\text{BD}}\to\infty$, unless the stress-energy-momentum tensor has a vanishing trace $T^{\text{(m)}}=0$ \cite{Brans Dicke infinite omega}, such as the matter content being radiation with $P_{\text{rad}}=\frac{1}{3}\rho_{\text{rad}}$ and $T^{\text{(m)}}_{\text{rad}}=-\rho_{\text{rad}}+3P_{\text{rad}}=0$.


\section{Energy-momentum conservation}\label{Sec Constraints from energy-momentum conservation}


In modified gravities with the generic Lagrangian density $\mathscr{L}=f(R,\mathcal{R}_i,\cdots)$, where $\mathcal{R}_i=\mathcal{R}_i \big(g_{\alpha\beta}\,,R_{\alpha\mu\beta\nu}\,,\nabla_\gamma R_{\alpha\mu\beta\nu}\,,\cdots\,,$ $
\nabla_{\gamma_ 1}\!\nabla_{\gamma_ 2}\ldots\nabla_{\gamma_ n} R_{\alpha\mu\beta\nu}\big)$ and the ``$\cdots$'' in $\mathscr{L}=f$ refer to arbitrary curvature invariants beyond the Ricci scalar, the energy-momentum conservation is naturally guaranteed by Noether's law or the generalized contracted Bianchi identities \cite{BDfR equivalence Generalized Bianchi}
\begin{equation}\label{GBIs}
\nabla^\mu \,\left( \frac{1}{\sqrt{-g}} \,\frac{\delta\, \left[\!\!\sqrt{-g}\,f(R,\mathcal{R}_i,\cdots)\right]}{\delta g^{\mu\nu}}\right)=\,0,
\end{equation}
which can be expanded into
\begin{equation}
f_R R_{\mu\nu}+\sum f_{\mathcal{R}_i}  \mathcal{R}_{\mu\nu}^{(i)} -\frac{1}{2}f(R,\mathcal{R}_i,\cdots) \,g_{\mu\nu}=0,
\end{equation}
where $f_R \coloneqq \partial f(R,\mathcal{R}_i,\cdots)/\partial R$, $f_{\mathcal{R}_i} \coloneqq \partial  f(R,\mathcal{R}_i,\cdots)/\partial \mathcal{R}_i$, and $ \mathcal{R}_{\mu\nu}^{(i)}\cong  \left(f_{\mathcal{R}_i} \delta \mathcal{R}_i \right)/\delta g^{\mu\nu}$.
However, in the more generic situations of scalar-tensor-type gravities with $\mathscr{L}=f(\phiup, R,\mathcal{R}_i,\cdots)$ $+\varpi\left(\phiup\,, \nabla_\alpha\phiup\nabla^\alpha\phiup\right)$ where nonminimal couplings between the scalar fields and the curvature invariants are involved, such as the LBD proposal under discussion, the conservation problem is more complicated than pure tensorial gravity.

Now let's get back to the LBD field equation (\ref{BDLL full 1}).
By the coordinate invariance or the diffeomorphism invariance of the matter action $\mathcal{S}_m$ in which $\mathscr{L}_m$ is neither coupled with the curvature invariants nor the scalar field $\phiup$, naturally we have the energy-momentum conservation $\nabla^\mu T_{\mu\nu}^{\text{(m)}}=0$ for the matter content. Thus, the covariant derivative of the left hand side of Eq.(\ref{BDLL full 1}) should also vanish. With the Bianchi identity $\nabla^\mu  \left(R_{\mu\nu} -\frac{1}{2}  R  g_{\mu\nu}\right)=0$ and the third-order-derivative commutator $ \left(\nabla_\nu\Box -\Box\nabla_\nu \right) \phiup=-R_{\mu\nu}\nabla^\mu\phiup$, it follows that
\begin{equation}\label{conservation constraint step 1}
\begin{split}
\nabla^\mu\left[\phiup \left(R_{\mu\nu} -\frac{1}{2}  R  g_{\mu\nu}\right)
+ \left(g_{\mu\nu}\Box -\nabla_\mu \nabla_\nu \right) \phiup \right]=-\frac{1}{2}R\,\nabla_\nu \phiup.
\end{split}
\end{equation}
Moreover, for the scalar field, we have
\begin{eqnarray}
&&\nabla^\mu\left[-\frac{\omega_{\text{L}}}{\phiup}\left(\nabla_\mu \phiup \nabla_\nu \phiup-\frac{1}{2}g_{\mu\nu}\nabla_\alpha \phiup \nabla^\alpha\phiup\right) \right] \nonumber\\
&&\,=\frac{1}{2}\nabla_\nu \phiup\cdot\left(\frac{\omega_{\text{L}}}{\phiup^2} \nabla_\alpha \phiup \nabla^\alpha\phiup- \frac{2\omega_{\text{L}}}{\phiup}\Box\phiup\right)\label{conservation constraint step 2}\\
&&\,=\frac{1}{2}\nabla_\nu \phiup\cdot\left(R+\frac{a}{\sqrt{-g}} {}^*RR +b \mathcal{G}  \right),\nonumber
\end{eqnarray}
where the kinematical wave equation (\ref{LBD wave kinematic}) has been employed.

For the Chern-Pontryagin and the Gauss-Bonnet parts in  Eq.(\ref{BDLL full 1}), consider the componential actions $\mathcal{S}_{\text{CP}}=\int d^4x\, \phiup {}^*RR$ and $\mathcal{S}_{\text{GB}}=\int d^4x\sqrt{-g} \phiup \mathcal{G}$.   Under an arbitrary infinitesimal coordinate transformation $x^\mu\mapsto x^\mu+\delta x^\mu$, where $\delta x^\mu= \xi^\mu$ is an infinitesimal vector field which vanishes on the boundary, so that the spacetime manifold is mapped onto itself. Then $\mathcal{S}_{\text{CP}}$ and  $\mathcal{S}_{\text{GB}}$ vary by
\begin{align}
&\delta \mathcal{S}_{\text{CP}}=-\int d^4x
\,\phiup\,\partial_\mu \left(\xi^\mu {}^*RR \right)
\cong \int d^4x {}^*RR\left(\partial_\mu\phiup \right)\xi^\mu,\label{conservation constraint step 5.6}\\
&\delta \mathcal{S}_{\text{GB}}=-\int d^4x \,\phiup\,\partial_\mu \left(\xi^\mu\sqrt{-g} \mathcal{G}\right)
\cong \int d^4x\sqrt{-g} \mathcal{G}\left(\partial_\mu\phiup \right)\xi^\mu. \label{conservation constraint step 6}
\end{align}
For the first step in Eqs.(\ref{conservation constraint step 5.6}) and Eqs.(\ref{conservation constraint step 6}), one should note that $x^\mu\mapsto x^\mu+\xi^\mu$ is a  particle/active transformation,
under which the dynamical
tensor fields transform, while the background scalar field $\phiup(x^\alpha)$ and the coordinate system parameterizing the spacetime manifold remain unchanged \cite{Diffeomorphism Explicit Spontaneous}.
On the other hand, the inverse metric transforms by $ g^{\mu\nu}\mapsto  g^{\mu\nu} +\delta g^{\mu\nu} $ with $\delta g^{\mu\nu}=-\pounds_{\vec\xi}g^{\mu\nu} =\nabla^\mu\xi^\nu+\nabla^\nu\xi^\mu$, and thus we have
\begin{eqnarray}
\delta \mathcal{S}_{\text{CP}}
=&&2\int d^4x  \sqrt{-g} \,H_{\mu\nu}^{\text{(CP)}}\, \nabla^\mu\xi^\nu \nonumber\\
\cong &-&2\int d^4x \sqrt{-g}\, \left(\nabla^\mu  H_{\mu\nu}^{\text{(CP)}} \right)\xi^\nu,\label{conservation constraint step 6.5}\\
\delta \mathcal{S}_{\text{GB}}
=&&2\int d^4x \sqrt{-g}\,  H_{\mu\nu}^{\text{(GB)}} \nabla^\mu\xi^\nu \nonumber\\
\cong& -&2\int d^4x\sqrt{-g}\, \left(\nabla^\mu  H_{\mu\nu}^{\text{(GB)}}\right)\xi^\nu.\label{conservation constraint step 7}
\end{eqnarray}
Comparing Eqs.(\ref{conservation constraint step 5.6}) with (\ref{conservation constraint step 6.5}), and Eqs.(\ref{conservation constraint step 6}) with (\ref{conservation constraint step 7}), we obtain the relations
\begin{equation}\label{conservation constraint step 4}
\nabla^\mu  H_{\mu\nu}^{\text{(CP)}}
=-\frac{1}{2}\frac{{}^*RR}{\sqrt{-g}}\cdot \partial_\nu \phiup\,,
\end{equation}
\begin{equation}\label{conservation constraint step 8}
\nabla^\mu  H_{\mu\nu}^{\text{(GB)}}
=-\frac{1}{2}\mathcal{G}\cdot\partial_\nu \phiup \,.
\end{equation}
Adding up Eqs.(\ref{conservation constraint step 1}), (\ref{conservation constraint step 2}), (\ref{conservation constraint step 4}), and (\ref{conservation constraint step 8}), one could find that the covariance divergence for the left hand side of the field equation  (\ref{BDLL full 1}) vanishes, which confirms the energy-momentum conservation in the LBD gravity.

Eqs.(\ref{conservation constraint step 4}) and (\ref{conservation constraint step 8}) for the nontrivial divergences of $H_{\mu\nu}^{\text{(CP)}}$ and $H_{\mu\nu}^{\text{(GB)}}$, by their derivation process, reflect the breakdown of diffeomorphism invariance for $\mathcal{S}_{\text{CP}}$ and $\mathcal{S}_{\text{GB}}$ in $\mathcal{S}_{\text{LBD}}$. They have clearly shown the influences of nonminimal $\phiup$-topology couplings to the covariant conservation, as opposed to the straightforward generalized Bianchi identities
\begin{equation}
\begin{split}
\nabla^\mu \left( \frac{1}{\sqrt{-g}}\frac{\delta {}^*RR}{\delta g^{\mu\nu}}\right)=0\;\;\;\mbox{and}\;\;\;
\nabla^\mu \left( \frac{1}{\sqrt{-g}}\frac{\delta (\sqrt{-g}\,\mathcal{G})}{\delta g^{\mu\nu}}\right)=0.
\end{split}
\end{equation}


\section{Conformal transformations}\label{Sec Conformal transformations}

The standard LBD action $\mathcal{S}_{\text{LBD}}$ in Eq.(\ref{LBD action 0}) can be transformed into different representations by conformal rescaling of the spacetime line element, which geometrically  preserves the angles between spacetime vectors and physically retains local causality structures.

\subsection{Dynamical Chern-Simons gravity}\label{Sec Conformal transformations into Chern-Simons gravity}

As a simplest example, consider the specialized  $\mathcal{S}_{\text{LBD}}$ in vacuum and for spacetimes of
negligible gravitational effects from the nonminimally $\phiup-$coupled Gauss-Bonnet term. With $\mathcal{S}_m=0$ and  $b=0$, Eq.(\ref{LBD action 0}) reduces to become
\begin{equation}\label{LBD action No G}
\mathcal{S}= \frac{1}{16\pi}\int d^4x\left[\sqrt{-g}\left( \phiup R
 -\frac{\omega_{\text{L}}}{\phiup}\nabla_\alpha \phiup
\nabla^\alpha\phiup \right)+a \phiup {}^*RR \right].
\end{equation}
For a pointwise scaling field $\Omega=\Omega(x^\alpha)>0$, we can rescale the metric $g_{\mu\nu}$ of the original frame into $\tilde{g}_{\mu\nu}$ via $\tilde{g}_{\mu\nu}=\Omega^2 g_{\mu\nu}$; it follows that $g_{\mu\nu}=\Omega^{-2} \tilde{g}_{\mu\nu}$,  $g^{\mu\nu}=\Omega^2 \tilde{g}^{\mu\nu}$, $\sqrt{-g}=\Omega^{-4}\sqrt{-\tilde{g}}$, and\footnote{Compared with $R=\Omega^2\left[\tilde{R}+6\tilde{\Box}  \Omega/\Omega-12\tilde{g}^{\alpha\beta}\partial_\alpha \Omega \partial_\beta \Omega /\Omega^2 \right]$, Eq.(\ref{conformal R}) best isolates pure-divergence terms and thus most simplifies the action once the coefficient of R is reset into unity. Moreover, by employing $\ln \Omega$ instead of $\Omega$, the transformations $R\to \tilde{R}$ becomes skew-symmetric to $\tilde{R}\to R$.}  \cite{Scalar Tensor Theory}
\begin{equation}\label{conformal R}
R=\Omega^{2}\left[\tilde{R}+6\tilde{\Box}(\ln \Omega)-6 \tilde{g}^{\alpha\beta}\partial_\alpha (\ln \Omega)\, \partial_\beta (\ln \Omega) \right].
\end{equation}
Hence, for the reduced LBD action Eq.(\ref{LBD action No G}), the conformal transformation
\begin{equation}\label{conformal rescaling}
 g_{\mu\nu} = \frac{1}{G\phiup} \cdot\tilde{g}_{\mu\nu}
\end{equation}
along with the redefinition of the scalar field \{$\vartheta=\vartheta(x^\alpha)$, $\phiup=\phiup(\vartheta)$\} lead to
\begin{eqnarray}\label{LBD action No G 2}
\mathcal{S}\cong \frac{1}{16\pi G}\int d^4x\, &&\left[\sqrt{-\tilde{g}}\left(\tilde{R}
 -\frac{2\omega_{\text{L}}+3}{2\phiup(\vartheta)^2} \left(\frac{d\phiup}{d\vartheta}\right)^2\tilde{\nabla}_\alpha \vartheta
\tilde{\nabla}^\alpha\vartheta \,\right)\right.\nonumber\\
&&\;+a\, \phiup(\vartheta) {}^*RR \Bigg],
\end{eqnarray}
where the scalar field $\vartheta$ no longer directly couples to the Ricci scalar $\tilde{R}$, and thus the $6\tilde{\Box}(\ln \Omega)$ component in Eq.(\ref{conformal R}) has been removed as it simply yields a  boundary term $6\int \partial_\alpha \left[\!\sqrt{-g} \partial^\alpha (\ln \Omega)\right] d^4x $ for the action. Also, Eq.(\ref{LBD action No G 2})
has utilized the fact that the $(1,3)$-type Weyl tensor $C^\alpha_{\;\;\beta\gamma\delta}$ and thus ${}^*RR={}^*CC=\widetilde{{}^*CC}=\widetilde{{}^*RR}$ are conformally invariant.
It is straightforward to observe from Eq.(\ref{LBD action No G 2}) that the kinetics of $\vartheta$ is canonical for $\omega_{\text{L}}>-3/2$, noncanonical for   $\omega_{\text{L}}<-3/2$, and nondynamical for  $\omega_{\text{L}}=-3/2$; here we are interested in the canonical case. For the specialization
\begin{equation}
d\vartheta=\pm\sqrt{2\omega_{\text{L}}+3}\,\frac{d\phiup}{\phiup},
\end{equation}
which integrates to yield
\begin{equation}
\vartheta=\pm\sqrt{2\omega_{\text{L}}+3}\, \ln \frac{\phiup}{\phiup_0},
\end{equation}
where $\phiup_0$ is an integration constant,
or inversely
\begin{equation}
\phiup=\phiup_0\exp \left(\pm\frac{\vartheta}{\sqrt{2\omega_{\text{L}}+3}}\right),
\end{equation}
the action Eq.(\ref{LBD action No G 3}) finally becomes
\begin{equation}\label{LBD action No G 3}
\begin{split}
\mathcal{S}= \frac{1}{16\pi G}\int d^4x \left[\sqrt{-\tilde{g}}\left(\tilde{R}
 -\frac{1}{2} \tilde{\nabla}_\alpha \vartheta
\tilde{\nabla}^\alpha\vartheta \right)\right.&\\
+a \phiup_0\exp \left(\pm\frac{\vartheta}{\sqrt{2\omega_{\text{L}}+3}}\right) \widetilde{{}^*RR} \Bigg]&.
\end{split}
\end{equation}
Hence, the conformal rescaling $g_{\mu\nu} = \tilde{g}_{\mu\nu}/G\phiup$ along with the new scalar field $\vartheta(x^\alpha)$ recast  the reduced LBD action Eq.(\ref{LBD action No G}) into Eq.(\ref{LBD action No G 3}), which is an action for the dynamical Chern-Simons gravity \cite{Chern-Simons 2}, though the nonminimal $\vartheta$--$\widetilde{{}^*RR}$ coupling is slightly more complicated than the straightforward $\vartheta\widetilde{{}^*RR}$ as in the popular Chern-Simons literature. Moreover, the conformal invariance of ${}^*RR$ guarantees that the effect of $\int d^4x\,\phiup {}^*RR$ could never be removed by conformal transformations.

Note that the matter action $\mathcal{S}_m(g_{\mu\nu}, \psi_m)$ would be transformed into $\mathcal{S}_m(\tilde{g}_{\mu\nu}/G\phiup, \psi_m)$ (in general $S_m$ does not contain derivatives of the metric tensor \cite{MTW Gravitation}), which are different in the $\phiup$--$\mathcal{S}_m$ or $\phiup$--$\mathscr{L}_m$ couplings; consequently $T_{\mu\nu}^{\text{(m)}}$ fails to be conformally invariant unless it is traceless $T^{\text{(m)}}=0$ \cite{Scalar Tensor Theory}. This is why we focus on the vacuum situation.


\subsection{Generalized Gauss-Bonnet dark energy}\label{Sec Conformal transformations into Gauss-Bonnet dark energy}

Similarly, in vacuum and for spacetimes of negligible Chern-Simons parity-violation effect, $\mathcal{S}_{\text{LBD}}$ reduces into
\begin{equation}\label{LBD action No CP}
\mathcal{S}=\frac{1}{16\pi}\int d^4x\left[\sqrt{-g}\left( \phiup R + b\phiup \mathcal{G}
 -\frac{\omega_{\text{L}}}{\phiup}\nabla_\alpha \phiup
\nabla^\alpha\phiup \right)\right].
\end{equation}
Under the local rescaling $g_{\mu\nu}\mapsto\tilde{g}_{\mu\nu}=\Omega^2 g_{\mu\nu}$ for the metric, the Gauss-Bonnet scalar satisfies \cite{Conformal transformations Gauss-Bonnet}
\begin{widetext}
\begin{equation}\label{Conformal GB}
\begin{split}
\mathcal{G}=\Omega^4 \Bigg\{\widetilde{\mathcal{G}}&  - 8\tilde{R}^{\alpha\beta}\tilde{\nabla}_\alpha\tilde{\nabla}_\beta (\ln \Omega)-8 \tilde{R}^{\alpha\beta}\tilde{\nabla}_\alpha (\ln \Omega)\tilde{\nabla}_\beta (\ln \Omega)+4\tilde{R}\,\tilde{\Box}(\ln \Omega)+8 \left[\tilde{\Box}(\ln \Omega)\right]^2
- 8 \tilde{\Box}(\ln \Omega) \cdot \tilde{\nabla}_\alpha (\ln \Omega)\tilde{\nabla}^\alpha (\ln \Omega)\\
&- 8\tilde{\nabla}_\alpha\tilde{\nabla}_\beta (\ln \Omega)\cdot
\tilde{\nabla}^\alpha\tilde{\nabla}^\beta (\ln \Omega)
- 16  \tilde{\nabla}_\alpha\tilde{\nabla}_\beta (\ln \Omega)  \cdot\tilde{\nabla}^\alpha (\ln \Omega)\tilde{\nabla}^\beta (\ln \Omega)\Bigg\}.
\end{split}
\end{equation}
\end{widetext}
Set the factor of conformal transformation to be
$\Omega=\sqrt{G\phiup}$ so that the Ricci scalar decouples from the scalar field, and redefine the scalar field via $\phiup(x^\alpha)\mapsto\varphi(x^\alpha)=\sqrt{2\omega_{\text{L}}+3}\, \ln \frac{\phiup}{\phiup_0}$ or equivalently $\phiup=\phiup_0\exp \left(\frac{\varphi}{\sqrt{2\omega_{\text{L}}+3}}\right)$; then it follows that
\begin{equation}\label{Conformal GB 2}
\begin{split}
\ln \Omega&=\frac{1}{2}\ln \phiup+\frac{1}{2}\ln G\\
&=\frac{1}{2}\frac{\varphi}{\sqrt{2\omega_{\text{L}}+3}}+\frac{1}{2}\ln\phiup_0 +\frac{1}{2}\ln G\,.
\end{split}
\end{equation}
With $\{\ln\phiup_0 ,\ln G\}$ being constants, substitution of  Eq.(\ref{Conformal GB 2})
 into Eq.(\ref{Conformal GB}) yields
\begin{equation}\label{Conformal GB 3}
\begin{split}
\sqrt{-g}\,\mathcal{G}=\sqrt{-\tilde{g}}\,\left( \widetilde{\mathcal{G}} + \frac{ \mathcal{K}(\tilde{\nabla}\varphi)}{\sqrt{2\omega_{\text{L}}+3}}\right)\quad\mbox{and}
\end{split}
\end{equation}
\begin{equation}\label{Conformal GB 4}
\begin{split}
\mathcal{K}(\tilde{\nabla}\varphi) = \,&- 2\tilde{R}^{\alpha\beta}\left(2\tilde{\nabla}_\alpha\tilde{\nabla}_\beta \varphi
 + \tilde{\nabla}_\alpha \varphi\tilde{\nabla}_\beta \varphi\right)\\
&+\tilde{\Box}\varphi\cdot \left(2\tilde{R}+2\tilde{\Box}\varphi - \tilde{\nabla}_\alpha \varphi\tilde{\nabla}^\alpha \varphi\right)\\
 &- 2\tilde{\nabla}_\alpha\tilde{\nabla}_\beta \varphi\cdot \left(
\tilde{\nabla}^\alpha\tilde{\nabla}^\beta \varphi
+\tilde{\nabla}^\alpha \varphi\tilde{\nabla}^\beta \varphi\right).
\end{split}
\end{equation}
Here one can observe that since the coefficient $\Omega^{-4}$ in $\sqrt{-g}=\Omega^{-4}\sqrt{-g}$ exactly neutralizes the  $\Omega^4$ in Eq.(\ref{Conformal GB}), the nonminimally $\phiup$-coupled Gauss-Bonnet effect $\int d^4x \sqrt{-g}\phiup\mathcal{G}$ could never be canceled by a conformal rescaling $g_{\mu\nu}=\Omega^{-2} \tilde{g}_{\mu\nu}$. Hence, the reduced LBD action Eq.(\ref{LBD action No CP}) is finally transformed into
\begin{equation}
\begin{split}
\mathcal{S}=&\frac{1}{16\pi G}\int d^4x \sqrt{-\tilde{g}}\,\Bigg\{\tilde{R}-\frac{1}{2}\tilde{\nabla}_\alpha\varphi \tilde{\nabla}^\alpha\varphi\, + \\
&b \phiup_0\exp \left(\frac{\varphi}{\sqrt{2\omega_{\text{L}}+3}}\right)
\left[\tilde{\mathcal{G}} + \mathcal{K}(\tilde{\nabla}\varphi)\right]\Bigg\},
\end{split}
\end{equation}
which generalizes the canonical Gauss-Bonnet  dark energy in vacuum $\mathcal{S}=\frac{1}{16\pi G}\int d^4x \sqrt{-g}\,\left(R-\frac{1}{2}\nabla_\alpha\varphi \nabla^\alpha\varphi + f(\varphi)\mathcal{G}\right)$ \cite{GaussBonnet dark energy} by the Horndeski-like \cite{Horndeski} or Galileon-like \cite{Galileon} kinetics in $\mathcal{K}(\tilde{\nabla}\varphi)$ for the scalar field.

Note that in the two examples just above, because of the nonminimal coupling to the scalar field $\phiup(x^\alpha)$, negligible Gauss-Bonnet effect does not imply a zero Euler characteristic $\chiup(\mathcal{M})= \frac{1}{32\pi^2} \int\!\sqrt{-g}\,\mathcal{G} d^4x =0$ for the spacetime, and similarly, negligibility of the Chern-Simons effect does not indicate a vanishing instanton number $\int{}^*RR \,d^4x =0$, either.

Also, for the actions of the Chern-Simons gravity and the Gauss-Bonnet dark energy in the Jordan frame, in which a scalar field is respectively coupled to ${}^*RR$ and $\mathcal{G}$, we cannot help but ask that why the scalar field is not simultaneously coupled to the Ricci scalar? We have previously seen from Eq.(\ref{GBIs}) that all algebraic and differential Riemannian invariants stand equal in front of the generalized Bianchi identities, so are there any good reasons for the scalar field to discriminate among different curvature invariants? We hope that the LBD gravity help release this tension (at least in empty spacetimes), as the scalar field $\phiup$ indiscriminately couples to all the LBD invariants $\{R,{}^*RR,\mathcal{G}\}$, and the LBD gravity takes the Chern-Simons gravity and the Gauss-Bonnet dark energy as its reduced representations in the Einstein frame.


\section{Cosmological applications}\label{Cosmological applications}

Having extensively discussed the theoretical structures of the LBD gravity, in this section we will apply this theory to the Friedman-Robertson-Walker (FRW) Universe and investigate the possibility to realize the late-time cosmic acceleration.

\subsection{Generalized Friedmann and Klein-Gordon equations}
The field equation (\ref{BDLL full 1}) can be recast into a GR form,
\begin{equation}\label{BDLL full 3}
\begin{split}
 R_{\mu\nu} -\frac{1}{2}  R  g_{\mu\nu} = \kappa^2 G_{\text{eff}} \left(T_{\mu\nu}^{\text{(m)}}+T_{\mu\nu}^{(\phiup)}
+T_{\mu\nu}^{\text{(CP)}}+T_{\mu\nu}^{\text{(GB)}}\right),
\end{split}
\end{equation}
where $\kappa^2=8\pi$, and $G_{\text{eff}}=1/\phiup$
denotes the effective gravitational coupling strength. $T_{\mu\nu}^{\text{(m)}}+T_{\mu\nu}^{(\phiup)}
+T_{\mu\nu}^{\text{(CP)}}+T_{\mu\nu}^{\text{(GB)}}\eqqcolon T_{\mu\nu}^{\text{(eff)}}$
compromises the total effective stress-energy-momentum tensor, with
\begin{eqnarray}\label{Tmunu CP GB}
&& \kappa^2 T_{\mu\nu}^{\text{(CP)}}=-a  H_{\mu\nu}^{\text{(CP)}}\;\;\;,\;\;\;  \kappa^2 T_{\mu\nu}^{\text{(GB)}}=-b H_{\mu\nu}^{\text{(GB)}},\;\;\mbox{and}\\
\kappa^2 &&T_{\mu\nu}^{(\phiup)}= \left(\nabla_\mu \nabla_\nu- g_{\mu\nu}\Box\right)\phiup+\frac{\omega_{\text{L}}}{\phiup}\left(\nabla_\mu \phiup \nabla_\nu \phiup-\frac{1}{2}g_{\mu\nu}\nabla_\alpha \phiup \nabla^\alpha\phiup\right)\nonumber.
\end{eqnarray}
Note that besides the effects of the source term $-\frac{\omega_{\text{L}}}{\phiup}\nabla_\alpha \phiup \nabla^\alpha\phiup$ in $\mathscr{L}_{\text{LBD}}$ via $\delta\left(-\sqrt{-g}\,\frac{\omega_{\text{L}}}{\phiup}\nabla_\alpha \phiup \nabla^\alpha\phiup\right)/\delta g^{\mu\nu}$, the $\left(\nabla_\mu \nabla_\nu- g_{\mu\nu}\Box\right)\phiup$ part from  $\delta\left(\sqrt{-g}\phiup R\right)/\delta g^{\mu\nu}$ is also packed into $T_{\mu\nu}^{(\phiup)}$. Moreover, with the four distinct components of $T_{\mu\nu}^{\text{(eff)}}$ sharing the same gravitational strength $1/\phiup$,
Eq.(\ref{BDLL full 3}) implicitly respects the equivalence principle that the gravitational interaction is independent of the internal structures and compositions of a test body or self-gravitating object \cite{Equivalence principle}.

For the FRW metric of the flat Universe with a vanishing spatial curvature index,
\begin{equation} \label{Flat FRW}
ds^2=-dt^2+a(t)^2\sum_{i=1}^3 \left(dx^i\right)^2,
\end{equation}
${}^*RR=0$ due to the maximal spatial symmetry, while the Ricci and Gauss-Bonnet scalars are respectively
\begin{equation}
\begin{split}
&R=6 \frac{a\ddot a+\dot{a}^2}{a^2} =6\left(\dot H+2H^2\right)\\
&\mathcal{G}=24\frac{\dot{a}^2\ddot{a}}{a^3}=24H^2\left(\dot H+H^2\right),
\end{split}
\end{equation}
where overdot denotes the derivative over the cosmic comoving time, and $H\coloneqq \dot{a}/a$ represents the time-dependent Hubble parameter. Thus, an accelerated/decelerated flat Universe has a positive/negative Euler-Poincar\'e topological density. With a perfect-fluid form $T^{\mu}_{\;\;\nu}=\text{diag}\left[-\rho,P,P,P\right]$ assumed for each component in $T_{\mu\nu}^{\text{(eff)}}$ [in consistency with the metric signature $(-,+++)$], the cosmic expansion satisfies the generalized Friedmann equations
\begin{equation}\label{LBD Friedmann 1}
\begin{split}
\hspace{-1cm}\left(\frac{\dot{a}}{a}\right)^2\,&=\frac{1}{3\phiup} \left(\kappa^2\rho_m -3H\dot\phiup+\frac{\omega_{\text{L}}}{2\phiup}\dot{\phiup}^2 -12bH^3\dot{\phiup}\right)\,,
\end{split}
\end{equation}
\begin{equation}\label{LBD Friedmann 2}
\begin{split}
\frac{\ddot{a}}{a}= -\frac{1}{6\phiup}\Bigg\{&\kappa^2\left(\rho_m+3P_m \right)
+3\ddot{\phiup}+3H\dot\phiup+\frac{2\omega_{\text{L}}}{\phiup}\dot{\phiup}^2\\
&+12b H^2 \ddot{\phiup}+ 12b\left(2\dot H +H^2\right)H\dot{\phiup}
 \Bigg\}\,,
\end{split}
\end{equation}
where $ T_{\mu\nu}^{\text{(CP)}}=0$ for FRW.
Moreover, the kinematical wave equation (\ref{LBD wave kinematic})
and the dynamical wave equation (\ref{LBD wave dynamical}) respectively lead to
\begin{equation} \label{LBD wave FRW kinematic}
\frac{2\omega_{\text{L}} }{\phiup} \left(\ddot\phiup+3H\dot\phiup\right)
= 6 \frac{a\ddot a+\dot{a}^2}{a^2}+\frac{ \omega_{\text{L}}}{\phiup^2}
+\ddot{\phiup} +24b H^2 \frac{\ddot{a}}{a},
\end{equation}
\begin{equation}\label{LBD wave FRW dynamical}
\begin{split}
\bigg(&2\omega_{\text{L}} +3+12b\frac{a\ddot a+\dot{a}^2}{a^2}\bigg)\left(\ddot\phiup+3H\dot\phiup\right) = 24b H^2\frac{\ddot{a}}{a}\phiup\\
& - 8\pi \left(3P_m-\rho_m\right)
+ 12 b\left( \frac{\ddot{a}}{a}\ddot{\phiup} + \frac{a\ddot{a}+2\dot{a}^2}{a^2} H\dot{\phiup}  \right).
\end{split}
\end{equation}
In principle, one could understand the evolutions of the scale factor $a(t)$ and the homogeneous scalar field $\phiup(t)$ by (probably numerically) solving Eqs.(\ref{LBD Friedmann 1})-(\ref{LBD wave FRW dynamical}). However the solutions will be complicated, so we will  start with some solution ansatz for \{$a(t)$,  $\phiup(t)$\}, which are easier to work with.


\subsection{Cosmic acceleration in the late-time approximation}

The physical matter satisfies the continuity equation
\begin{equation}
\dot{\rho}_m+3H(\rho_m+P_m)=0,
\end{equation}
and for pressureless dust $P_m=0$, it integrates to yield
\begin{equation}\label{cosmology solu 1}
\rho_m=\rho^{(m)}_0a^{-3}=\frac{\rho^{(m)}_0}{a_0^3 }t^{-3\betaup},
\end{equation}
where we have assumed a power-law scale factor
\begin{equation}\label{cosmology solu 2}
a=a_0 t^{\betaup}\quad\mbox{with}\;\;\betaup>1.
\end{equation}
Here \{$a_0$, $\betaup$\} are constants, and $\betaup>1$ so that $\ddot{a}>0$.
Similarly, we also take a power-law ansatz for the scalar field,
\begin{equation}\label{cosmology solu 4}
\phiup=\phiup_0 t^{\gammaup}.
\end{equation}
Based on Eqs.(\ref{cosmology solu 1})-(\ref{cosmology solu 4}), the dynamical wave equation (\ref{LBD wave dynamical}) with
$T^{\text{(m)}}=-\rho_m$ for dust yields
\begin{equation}
\begin{split}
\gammaup\left(2\omega_{\text{L}} +3\right)\left( 3\betaup-1+  \gammaup\right) =  \frac{\kappa^2\rho^{\text{(m)}}_0}{\phiup_0 a_0^3} t^{2-3\betaup-\gammaup}\\
 +24 b\frac{\betaup^3(\betaup-1)}{t^2}
-12 b \betaup^2\gammaup\frac{\left(3\betaup-3+\gammaup\right)}{t^2},
\end{split}
\end{equation}
and in the late-time (large $t$) approximation it reduces to
\begin{equation}
\gammaup\left( \gammaup+ 3\betaup-1\right)= \frac{1}{\phiup_0 }\frac{\kappa^2\rho^{\text{(m)}}_0 }{ a_0^3 \left(2\omega_{\text{L}} +3\right)} t^{2-3\betaup-\gammaup},
\end{equation}
which can be satisfied by
\begin{equation}\label{cosmology solu 5}
\gammaup=2-3\betaup\;\;\;\mbox{and}\;\;\;\phiup_0 = \frac{\kappa^2\rho^{\text{(m)}}_0 }{ a_0^3 \left(2\omega_{\text{L}} +3\right)\left(2-3\betaup\right)}.
\end{equation}
Moreover, the first Friedmann equation (\ref{LBD Friedmann 1}) leads to
\begin{equation}
3\betaup^2=\kappa^2\frac{\rho^{(m)}_0}{a_0^3\phiup_0 } t^{2-3\betaup-\gammaup}
-3\betaup\gammaup+\frac{\omega_{\text{L}}}{2}\gammaup^2 -12b\frac{\betaup^3\gammaup}{t^2},
\end{equation}
and with Eq.(\ref{cosmology solu 5}), in the late-time approximation it becomes
\begin{equation}\label{beta omega 1st Friedmann}
3\betaup^2=(2-3\betaup)\left(2\omega_{\text{L}} +3\right)
-3\betaup(2-3\betaup)+\frac{\omega_{\text{L}}}{2}(2-3\betaup)^2.
\end{equation}
For $\betaup=2$, Eq.(\ref{beta omega 1st Friedmann}) trivially holds for an arbitrary  $\omega_{\text{L}}$, while for $\betaup\neq 2$, we have $\betaup$ in terms of  $\omega_{\text{L}}$ via
\begin{equation}\label{cosmology solu 6}
\betaup =  \frac{2\left(\omega_{\text{L}}+1\right)}{3\omega_{\text{L}}+4}.
\end{equation}
Note that Eqs.(\ref{cosmology solu 5}) and (\ref{cosmology solu 6}) reuire $\omega_{\text{L}}\neq-4/3$, $\omega_{\text{L}}\neq -3/2$ ($\betaup\neq 2$), and $\beta\neq 2/3$; they are simply consequences of the power-law-solution ansatz and the late-time approximations rather than universal constraints on $\omega_{\text{L}}$, and according to Eq.(\ref{cosmology solu 6}), the last condition $\beta\neq 2/3$ trivially holds with  $\beta\to 2/3$ for $\omega_{\text{L}}\to \infty$. As a consistency test,
the kinematical equation (\ref{LBD wave kinematic}) yields
\begin{equation}\label{cosmology solu 7}
\omega_{\text{L}} \left(\frac{1}{2}\gamma^2-\gamma+3\beta\gamma\right)\,=
3\beta(2\beta-1) + 12 b\frac{\beta^3(\beta-1)}{t^2}
\end{equation}
with the late-time approximation
\begin{equation}
\omega_{\text{L}} \left(\frac{1}{2}\gamma^2-\gamma+3\beta\gamma\right)\,=
3\beta(2\beta-1),
\end{equation}
which holds for Eqs.(\ref{cosmology solu 5}) and (\ref{cosmology solu 6}).
Substituting Eqs.(\ref{cosmology solu 1}), (\ref{cosmology solu 2}), (\ref{cosmology solu 4}), (\ref{cosmology solu 5}) and (\ref{cosmology solu 6}) into the second Friedmann equation (\ref{LBD Friedmann 2}), we obtain
\begin{equation}\label{Cosmo acce 1}
\frac{\ddot{a}}{a}=-\frac{2(\omega_{\text{L}}+1)(\omega_{\text{L}}+2)}{(3\omega_{\text{L}}+4)^2} t^{-2},
\end{equation}
and the deceleration parameter reads
\begin{equation}\label{Cosmo acce 2}
\begin{split}
q\,\coloneqq-\frac{\ddot{a}a}{\dot{a}^2}=\frac{1}{2}\left(1+3\frac{P_{\text{eff}}}{\rho_{\text{eff}}}\right)=\frac{\omega_{\text{L}}+2}{2(\omega_{\text{L}}+1)}.
\end{split}
\end{equation}
Eqs.(\ref{Cosmo acce 1}) and (\ref{Cosmo acce 2}) clearly indicate that the late-time acceleration could be realized for $-2<\omega_{\text{L}}<-1$ ($\omega_{\text{L}}\neq -4/3$, $\omega_{\text{L}}\neq -3/2$), although this domain of $\omega_{\text{L}}$ makes the kinetics of the scalar field noncanonical.


\section{Lovelock-scalar-tensor gravity}\label{Lovelock-scalar-tensor gravity}


\subsection{From LBD to Lovelock-scalar-tensor gravity}

The LBD gravity can be generalized into the Lovelock-scalar-tensor (LST) gravity with the action
\begin{equation}\label{Scalar-tensor-Lanczos-Lovelock action I}
\begin{split}
\mathcal{S}_{\text{LST}}=&\int d^4x \sqrt{-g}\,\mathscr{L}_{\text{LST}}+\mathcal{S}_m\quad\mbox{and}\\
\mathscr{L}_{\text{LST}}=&\frac{1}{16\pi G}\left(f_1(\phiup) R+ f_2(\phiup)\frac{{}^*RR}{\sqrt{-g}} + f_3 (\phiup)  \mathcal{G} \right.\\
&-\frac{\omega(\phiup)}{\phiup} \nabla_\alpha \phiup \nabla^\alpha\phiup -V(\phiup)\Bigg)\,,
\end{split}
\end{equation}
where $\left\{f_i(\phiup), \omega(\phiup)\right\}$ are generic functions of the scalar field, and $V(\phiup)$ is the self-interaction potential. Note that this time Newton's constant $G$ is included in the overall coefficient $1/16\pi G$ of $\mathscr{L}_{\text{LST}}$, as is the case of the ordinary scalar-tensor gravity.
The gravitational field equation is
\begin{equation}\label{LST full 1}
\begin{split}
&f_1(\phiup) \left(R_{\mu\nu} -\frac{1}{2}  R  g_{\mu\nu}\right)+ \left(g_{\mu\nu}\Box -\nabla_\mu \nabla_\nu \right)f_1(\phiup)\\
&-\frac{\omega(\phiup)}{\phiup}\left(\nabla_\mu \phiup \nabla_\nu \phiup-\frac{1}{2}g_{\mu\nu}\nabla_\alpha \phiup \nabla^\alpha\phiup\right) \\
& +\frac{1}{2}V(\phiup)g_{\mu\nu}+  \widetilde{H}_{\mu\nu}^{\text{(CP)}} + \widetilde{H}_{\mu\nu}^{\text{(GB)}} =8\pi  T_{\mu\nu}^{\text{(m)}},
\end{split}
\end{equation}
where $\widetilde{H}_{\mu\nu}^{\text{(CP)}}$ denotes the contribution from
$f_2(\phiup) {}^*RR$,
\begin{equation}
\begin{split}
\sqrt{-g}\,\widetilde{H}_{\mu\nu}^{\text{(CP)}}= &2\partial^\xi f_2(\phiup)\cdot
\left( \epsilon_{\xi\mu\alpha\beta}\nabla^\alpha R^\beta_{\;\;\nu} +
\epsilon_{\xi\nu\alpha\beta}\nabla^\alpha R^\beta_{\;\;\mu} \right)\\
&+2\partial_\alpha \partial_\beta  f_2(\phiup)\cdot \left({}^*R^{\alpha\;\;\,\beta}_{\;\;\,\mu\;\;\nu} +{}^*R^{\alpha\;\;\,\beta}_{\;\;\,\nu\;\;\mu}\right)\,,
\end{split}
\end{equation}
and $\widetilde{H}_{\mu\nu}^{\text{(GB)}}$ attributes to the effect of $\sqrt{-g}f_3(\phiup)  \mathcal{G}$,
\begin{equation}
\begin{split}
\widetilde{H}_{\mu\nu}^{\text{(GB)}} =& 2R\left(g_{\mu\nu}\Box
-\nabla_\mu\!\nabla_\nu\right)  f_3(\phiup)-4R_{\mu\nu}\Box  f_3(\phiup)\,+\\
 &4R_{\mu}^{\;\;\,\alpha}\nabla_\alpha \nabla_{\nu} f_3(\phiup)   +4R_{\nu}^{\;\;\,\alpha}\nabla_\alpha \nabla_{\mu}  f_3(\phiup) \,- \\
 &4g_{\mu\nu}  R^{\alpha\beta}\nabla_\alpha\!\nabla_\beta f_3(\phiup)  +4 R_{\alpha\mu \beta\nu} \nabla^\beta \nabla^\alpha   f_3(\phiup) .
\end{split}
\end{equation}
It is straightforward to derive  the kinematical wave equation by $\delta\mathcal{S}_{\text{LST}}/\delta\phiup=0$, which along with the trace of Eq.(\ref{LST full 1}) could yield the dynamical wave equation, and they generalize the wave equations (\ref{LBD wave kinematic}, \ref{LBD wave dynamical}) in the LBD gravity. The wave equations however will not be listed here as the interest of this section is only the field equation $\delta\mathcal{S}/\delta g^{\mu\nu}=0$.


\subsection{Equivalence of LST with fourth-order gravities}

It is well known that the $f(R)$ gravity is equivalent to the nondynamical (i.e. $\omega_{\text{BD}}=0$) Brans-Dicke gravity \cite{BDfR equivalence Generalized Bianchi}, and such equivalence holds for the LBD gravity as well.
Consider the fourth-order modified gravity
\begin{equation}\label{LST MG equiv action}
\begin{split}
\mathscr{L}=&\frac{1}{16\pi G}\left[ f(R,  \mathcal{G}) +h\left(\frac{^*RR}{\sqrt{-g}}\right)\right],
\end{split}
\end{equation}
for which the field equation is
\begin{equation}\label{LST MG equiv}
\begin{split}
&f_R  R_{\mu\nu}+ \left(g_{\mu\nu}\Box -\nabla_\mu \nabla_\nu \right)f_R -\frac{1}{2}f(R,\mathcal{G})g_{\mu\nu}\\
&+  \mathcal{H}_{\mu\nu}^{\text{(CP)}} + \mathcal{H}_{\mu\nu}^{\text{(GB)}} =8\pi  T_{\mu\nu}^{\text{(m)}},
\end{split}
\end{equation}
where
\begin{equation}
\begin{split}
\sqrt{-g}\,\mathcal{H}_{\mu\nu}^{\text{(CP)}}= &2\partial^\xi h_{{}^*RR}\cdot
\left( \epsilon_{\xi\mu\alpha\beta}\nabla^\alpha R^\beta_{\;\;\nu} +
\epsilon_{\xi\nu\alpha\beta}\nabla^\alpha R^\beta_{\;\;\mu} \right)\\
& +2\partial_\alpha \partial_\beta  h_{{}^*RR} \cdot \left({}^*R^{\alpha\;\;\,\beta}_{\;\;\,\mu\;\;\nu} +{}^*R^{\alpha\;\;\,\beta}_{\;\;\,\nu\;\;\mu}\right)\,,
\end{split}
\end{equation}
and
\begin{equation}
\begin{split}
\mathcal{H}_{\mu\nu}^{\text{(GB)}} =& 2R\left(g_{\mu\nu}\Box
-\nabla_\mu\!\nabla_\nu\right)  f_{\mathcal G}-4R_{\mu\nu}\Box  f_{\mathcal G}\,+\\
 &4R_{\mu}^{\;\;\,\alpha}\nabla_\alpha \nabla_{\nu} f_{\mathcal G}
 +4R_{\nu}^{\;\;\,\alpha}\nabla_\alpha \nabla_{\mu}  f_{\mathcal G}\, - \\
 &4g_{\mu\nu}  R^{\alpha\beta}\nabla_\alpha\!\nabla_\beta f_{\mathcal G}  +4 R_{\alpha\mu \beta\nu} \nabla^\beta \nabla^\alpha   f_{\mathcal G},
\end{split}
\end{equation}
with $f_R=f_R(R, \mathcal{G})=\partial f(R, \mathcal{G})/\partial R$, $f_{\mathcal{G}}=f_{\mathcal{G}}(R, \mathcal{G})=\partial f(R, \mathcal{G})/\partial \mathcal{G}$, and $h_{{}^*RR}=d h({}^*RR/\sqrt{-g})/d ({}^*RR/\sqrt{-g})$. 
For the nondynamical LST gravity with $\omega(\phiup)\equiv0$ in Eq.(\ref{LST full 1}), compare it with Eq.(\ref{LST MG equiv}) and at the level of the gravitational equation, one could find the equivalence
\begin{equation}\label{Equivalence LST}
\begin{split}
f_1(\phiup)=f_R\;\;,\;\; &f_3(\phiup)=f_{\mathcal{G}}  \;\;,\;\; f_2(\phiup)=h_{{}^*RR }, \\
V(\phiup)=-&f(R, \mathcal{G})+ f_R R\,.
\end{split}
\end{equation}
In the $V(\phiup)$ relation we have applied the replacement $f_1(\phiup)=f_R$, and note that  $V(\phiup)$ does not contain a $f_{\mathcal{G}} \mathcal{G}$ term which has been removed from $\mathcal{H}_{\mu\nu}^{\text{(GB)}}$ because of the Bach-Lanczos identity Eq.(\ref{GaussBonnet Bach-Lanczos identity II}).


\subsection{Partial equivalence for ``multi-scalar LBD gravity''}

Removing the $\omega_{\text L}$ term in Eq.(\ref{BDLL full 1}) and then comparing it with Eq.(\ref{LST MG equiv}), one could find that an equivalence between the nondynamical LBD gravity (now equipped  with an extra potential $-U(\phiup)$ in $\mathscr{L}_{\text{LBD}}$) and the $f(R,  \mathcal{G}) +h\left(\frac{{}^*RR}{\sqrt{-g}}\right)$ gravity would require $f_R=f_\mathcal{G}=\phiup=h_{{}^*RR}$,  and $U(\phiup)=-f(R,\mathcal{G})$ $+f_R R$. These conditions are so restrictive that the $f(R,  \mathcal{G}) +h\left(\frac{^*RR}{\sqrt{-g}}\right)$ gravity would totally lose its generality. Instead,
introduce three auxiliary fields $\{\chi_1,  \chi_2, \chi_3\}$
and consider the dynamically equivalent action
\begin{align}
\mathcal{S}=&\frac{1}{16\pi}\int d^4x \sqrt{-g}\,\bigg[ f(\chi_1,\chi_2)+f_{\chi_1}\cdot(R-\chi_1)+f_{\chi_2}\cdot(\mathcal{G}-\chi_2)\nonumber\\
&+h(\chi_3)+h_{\chi_3}\cdot
\left(\frac{^*RR}{\sqrt{-g}}-\chi_3\right)\bigg]+\mathcal{S}_m\,;
\end{align}
its variation with respect to $\chi_1$,  $\chi_2$, and $\chi_3$ separately yields the constraints
\begin{equation}\label{Chi Constraints}
\begin{split}
f_{\chi_1\chi_1}&(R-\chi_1)=0\,,\quad
f_{\chi_2\chi_2}(\mathcal{G}-\chi_2)=0\,,\\
&\;\text{and}\quad  h_{\chi_3\chi_3}\Big(\frac{^*RR}{\sqrt{-g}}-\chi_3\Big)=0\,,
\end{split}
\end{equation}
where $f_{\chi_j}\coloneqq\partial f(\chi_1,\chi_2)/\partial\chi_j$, $f_{\chi_j\chi_j}\coloneqq\partial^2 f(\chi_1,\chi_2)/\partial\chi_j^2$, $h_{\chi_3}\coloneqq \partial h(\chi_3)/\partial \chi_3$ and $h_{\chi_3\chi_3}\coloneqq \partial^2 f(\chi_3)/\partial\chi_3^2$. If $f_{\chi_1\chi_1}$,  $f_{\chi_2\chi_2}$ and $h_{\chi_3\chi_3}$ do not vanish identically, Eq.(\ref{Chi Constraints}) leads to $\chi_1= R$, $\chi_2=\mathcal{G}$ and $\chi_3=\frac{^*RR}{\sqrt{-g}}$. Redefining the fields $\{\chi_1,  \chi_2, \chi_3\}$ by
\begin{equation}\label{MultiScalar LBD I}
\begin{split}
\phiup=f_{\chi_1}
\,,\quad\psiup=f_{\chi_2}
\,,\quad\varphiup=h_{\chi_3}\\
\end{split}
\end{equation}
and setting
\begin{align}\label{MultiScalar LBD II}
V(\phiup,\psiup,\varphiup)=&\phiup\cdot R(\phiup,\psiup)+\psiup\cdot\mathcal{G}(\phiup,\psiup)
+\varphiup\cdot\frac{{}^*RR}{\sqrt{-g}}(\varphiup)\nonumber \\
&-f\Big(R(\phiup,\psiup),\mathcal{G}(\phiup,\psiup)\Big)
-h\left(\frac{{}^*RR}{\sqrt{-g}}(\varphiup)\right),
\end{align}
then  the $f(R,  \mathcal{G}) +h\left(\frac{{}^*RR}{\sqrt{-g}}\right)$ gravity is partially equivalent to the following ``multi-scalar LBD gravity'' carrying three nondynamical scalar fields
\begin{equation}\label{MultiScalar LBD III}
\mathscr{L}=\frac{1}{16\pi}\left( \phiup R
+\varphiup\frac{{}^*RR}{\sqrt{-g}} + \psiup\mathcal{G} -V(\phiup,\psiup,\varphiup)\right),
\end{equation}
where the coupling coefficients $\{a,b\}$ appearing in $\mathscr{L}_{\text{LBD}}$ have been absorbed into the scalar fields $\{\varphiup, \psiup\}$.
Also, by ``partially equivalent'' we mean that
Eq.(\ref{MultiScalar LBD II}) as is stands is only partially on-shell; to recover Eq.(\ref{LST MG equiv action}) from the multi-field action of Eq.(\ref{MultiScalar LBD III}), one would have to add extra Lagrange multipliers identifying the different fields, but this would break the exact equivalence between such modified Eq.(\ref{MultiScalar LBD III}) and Eq.(\ref{LST MG equiv action}).


%


\section{Conclusions and discussion}

The Hilbert-Einstein action $\mathcal{S}_{\text{HE}}$  and the Lovelock action $\mathcal{S}_{\text L}$ yield identical field equations and thus are observationally indistinguishable. However, the former takes the Brans-Dicke gravity as its scalar-tensor counterpart, while the latter's companion is the LBD gravity, and these two theories are different.

We have extensively studied the theoretical structures of the LBD gravity, including the gravitational and wave equations, the ordinary $\omega_{\text L}\to\infty$ limit that recovers GR,
the unusual $\omega_{\text L}\to\infty$ limit satisfying the topology balance condition Eq.(\ref{topological balance condition}) and thus departing from GR, the energy-momentum conservation, the conformal transformations into the dynamical Chern-Simons gravity  and the generalized Gauss-Bonnet dark energy, as well as the extensions to LST gravity with its equivalence to fourth-order modified gravity.

We have taken the opportunity of deriving the field equation to look deeper into the properties of the Chern-Pontryagin and Gauss-Bonnet topological invariants. Especially, for the $f(\phiup)\mathcal{G}$ Gauss-Bonnet dark energy as well as the $f(R,\mathcal{G})$ and $f(R,\mathcal{G}, \mathscr{L}_m)$ gravities,  the contributions of the generalized Gauss-Bonnet dependence could be simplified from the
popular form like Eq.(\ref{GB variation II}) into our form like Eq.(\ref{GB variation III}).

An important goal of alternative and modified gravities is to explain the accelerated expansion of the Universe, and we have applied the LBD theory to this problem, too. It turned out that the acceleration could be realized for $-2<\omega_{\text L}<-1$ under our solution ansatz. Note that our estimate of cosmic acceleration in Sec.~\ref{Cosmological applications} is not satisfactory. For example, the kinematical equation (\ref{cosmology solu 7}) clearly shows that because of the higher-order time derivative terms arising from the $\phiup \mathcal{G}$ dependence, the simplest solution ansatz \{$\phiup=\phiup_0 t^{\gammaup}$,  $a=a_0 t^{\betaup}$\} with \{$\betaup$=constant, $\gammaup$=constant\} are not compatible with each other unless the late-time approximation is imposed, while such approximations further lead to the behaviors analogous to the Brans-Dicke cosmology \cite{Brans-Dicke cosmic acceleration}.

Section~\ref{Cosmological applications} has shown that,
the effects from the parity-violating Chern-Pontryagin term $\phiup {}^*RR$ are ineffective for the FRW cosmology because of its spatial homogeneity and isotropy. However, it is believed that  $\phiup {}^*RR$ could have detectable consequences on leptogenesis and gravitational waves in the initial inflation epoch \cite{CP inflation} where $\phiup$ acts as the inflaton field. The inflation problem usually works with the slow-roll approximations $\ddot\phiup\ll\dot\phiup\ll H$ and requires the existence of a potential well $V(\phiup)$; thus, at least for the description of the initial inflation, the LBD gravity should be generalized to carry a potential:
\begin{equation}
\widehat{\mathscr{L}}_{\text{LBD}}=  \frac{1}{16\pi}\left[ \phiup \left(R
+a\frac{{}^*RR}{\sqrt{-g}} + b\mathcal{G}\right)
 -\frac{\omega_{\text{L}}}{\phiup}\nabla_\alpha
\phiup \nabla^\alpha\phiup-V(\phiup) \right],
\end{equation}
with $V(\phiup)=2\Lambda\phiup$ being the simplest possibility.

Our prospective studies aim to construct the complete history of cosmic expansion in LBD gravity [probably equipped with $V(\phiup)$], throughout the dominance of radiation, dust, and effective dark energy. Moreover, it is well known that primordial gravitational waves can trace back to the Planck era of the Universe and serve as one of the most practical and efficient tests for modified gravities, so it is very useful to find out whether the gravitational-wave polarizations carry different intensities in this gravity. There are also some other problems from the LBD gravity attracting our attention, such as its relation to the low-energy effective string theory. We will look for the answers in future.


\section*{Acknowledgement}

DWT is very grateful to Prof. Christian Cherubini (Universit\`{a} Campus Bio-Medico) for helpful discussion on the topological current for the Gauss-Bonnet invariant. Also, we are very grateful to the anonymous referees for constructive suggestions to improve the manuscript. This work was financially supported by the Natural Sciences and Engineering Research Council of Canada, under the grant 261429-2013.


\begin{thebibliography}{200}
\bibitem{Convention switch}

Switching from the sign convention in Ref.\cite{Chern-Simons 1} to the convention in our paper and Ref.\cite{MTW Gravitation}, the signs for the following quantities are changed: $g_{\mu\nu}$, $\Gamma_{\alpha\beta\gamma}$, $R^\alpha_{\;\;\,\mu\beta\nu}$, $R_{\mu\nu}=R^\alpha_{\;\;\,\mu\alpha\nu}$, $G_{\mu\nu}=R_{\mu\nu}-\frac{1}{2}Rg_{\mu\nu}$, $T_{\mu\nu}^{\text{(m)}}$, $\Lambda$ (cosmological constant)
while the following quantities take unchanging values: $\sqrt{-g}$, $\epsilon^{\alpha\beta\mu\nu}$, $\epsilon_{\alpha\beta\mu\nu}$, $\Gamma^\alpha_{\mu\nu}$,  $R_{\alpha\mu\beta\nu}$, $R=g^{\mu\nu}R_{\mu\nu}$. Moreover, Ref.\cite{Chern-Simons 1} defined ${}^*RR$ by $\Big(\epsilon_{\alpha\beta\mu\nu} R^{\mu\nu}_{\;\;\;\;\gamma\delta}\Big)\,R^{\alpha\beta\gamma\delta}
=  2{}^{*}R_{\alpha\beta\gamma\delta}\,R^{\alpha\beta\gamma\delta}
=-2{}^*RR$, differing from ours by a minus sign, as we choose to follow Lovelock's usage which is more popular and standard.
\end{thebibliography}
\end{document}